\documentclass[12pt]{article}

\usepackage{amsmath}
\usepackage{amssymb}
\allowdisplaybreaks
\usepackage{epsf}

\begin{document}
\baselineskip=14.8pt plus 0.2pt minus 0.1pt
\renewcommand{\theequation}{\thesection.\arabic{equation}}
\renewcommand{\thefootnote}{\fnsymbol{footnote}}
\makeatletter
\@addtoreset{equation}{section}
\makeatother

\newcommand{\VEV}[1]{\left\langle #1\right\rangle}
\newcommand{\wt}{\widetilde}
\newcommand{\wh}{\widehat}
\newcommand{\ol}{\overline}
\newcommand{\p}{\partial}
\newcommand{\nn}{\nonumber}
\newcommand{\tr}{\mathop{\rm tr}\nolimits}
\newcommand{\Tr}{\mathop{\rm Tr}\nolimits}
\newcommand{\diag}{\mathop{\rm diag}}
\newcommand{\cL}{{\mathcal L}}
\newcommand{\cO}{{\mathcal O}}
\newcommand{\cA}{{\mathcal A}}
\newcommand{\cF}{{\mathcal F}}
\newcommand{\cI}{{\mathcal I}}
\newcommand{\cD}{{\mathcal D}}
\newcommand{\abs}[1]{\left| #1\right|}
\newcommand{\drv}[2]{\frac{d #1}{d #2}}
\newcommand{\Drv}[2]{\frac{\p #1}{\p #2}}
\newcommand{\YM}{{\rm YM}}
\newcommand{\CS}{{\rm CS}}
\newcommand{\MeV}{{\rm MeV}}
\newcommand{\fm}{{\rm fm}}
\newcommand{\bra}[1]{\left\langle #1\right\vert}
\newcommand{\ket}[1]{\left\vert #1\right\rangle}
\newcommand{\braket}[2]{\langle #1\vert #2\rangle}
\newcommand{\cl}{{\rm cl}}
\newcommand{\Ah}{\widehat{A}}
\newcommand{\ds}{\displaystyle}
\newcommand{\bm}[1]{\boldsymbol{#1}}
\newcommand{\SCSnew}{S_{\rm CS}^{\rm new}}
\newcommand{\gtfn}{\zeta}
\newcommand{\Po}{{\rm P}}
\newcommand{\To}{{\rm T}}
\newcommand{\veps}{\varepsilon}
\newcommand{\bVEV}[1]{\bigl\langle #1\bigr\rangle}
\newcommand{\QB}{Q_{\rm B}}
\newcommand{\fJ}{J}
\newcommand{\ginst}{g_{\rm inst}}
\newcommand{\olPhi}{\ol{\Phi}}
\newcommand{\elmg}{{\rm em}}
\newcommand{\MKK}{M_{\rm KK}}
\newcommand{\rhost}{\rho_{\rm st}}
\newcommand{\ANW}{\mbox{ANW}}
\newcommand{\Exp}{\mbox{Exp}}
\newcommand{\regg}{\text{reg.\ gauge}}
\newcommand{\Debar}{$\overline{\mbox{D8}}$}
\newcommand{\ellN}{{\wt{\ell}_N}}
\newcommand{\rhoVEV}[1]{\VEV{#1}_{\!\rho}}

\begin{titlepage}

\title{
\hfill\parbox{4cm}
{\normalsize
KUNS-2129\\ YITP-08-11}\\
\vspace{1cm}
{\bf
Chiral currents and static properties of nucleons
in holographic QCD
}}

\author{
Hiroyuki {\sc Hata},$^{a}$\,\thanks{
{\tt hata@gauge.scphys.kyoto-u.ac.jp}}
\,\ Masaki {\sc Murata}$^{b}$\,\thanks{
{\tt masaki@yukawa.kyoto-u.ac.jp}}
\ and Shinichiro {\sc Yamato}$^{a}$\,\thanks{
{\tt yamato@gauge.scphys.kyoto-u.ac.jp}}
\\[7mm]
$^{a}${\it
Department of Physics, Kyoto University, Kyoto 606-8502, Japan
}\\[3mm]
$^{b}${\it Yukawa Institute for Theoretical Physics, Kyoto University}\\
{\it Kyoto 606-8502, Japan}
}
\date{{\normalsize March 2008}}
\maketitle

\begin{abstract}
\normalsize
We analyze static properties of nucleons in the two flavor holographic
QCD model of Sakai and Sugimoto described effectively by a
five-dimensional $U(2)$ Yang-Mills theory with the Chern-Simons term
on a curved background.
The baryons are represented in this model as a soliton, which at a
time slice is approximately the BPST instanton with a fixed size.
First, we construct a chiral current in four dimensions from the
Noether current of local gauge transformations which are non-vanishing
on the boundaries of the extra dimension. We examine this chiral
current for nucleons with quantized collective coordinates to
compute their charge distribution, charge radii, magnetic moments and
axial vector coupling. Most of the results are better close to the
experimental values than in the Skyrme model.
We discuss the problems of our chiral current; non-uniqueness of the
local gauge transformation for defining the current, and its
gauge-noninvariance.

\end{abstract}

\thispagestyle{empty}
\end{titlepage}

\section{Introduction}
\label{sec:Intro}

Holographic QCD of Sakai and Sugimoto \cite{SaSu1,SaSu2} has
attracted much attention recently as a phenomenologically and
theoretically interesting model.
This model is based on a configuration
of $N_f$ D8-\Debar-branes and $N_c$ D4-branes, and the D8-D4,
\Debar-D4 and D4-D4 open strings correspond to the left-handed quark,
the right-handed quark and the gluon, respectively.
Spontaneous chiral symmetry breaking is implemented in this model by
the fact the D8- and \Debar-branes are smoothly connected.
Using holographic approximation to replace the D4-branes with the
near-horizon geometry of the corresponding SUGRA solution, the low
energy effective action of the D8-branes is expressed as a
five-dimensional Yang-Mills (YM) theory with the Chern-Simons (CS)
term on a curved background.
(We call this five-dimensional YM+CS theory simply the SS-model
hereafter.)
Mode-expanding the fields with respect to the coordinate $z$ of
the extra fifth dimension, we obtain a four-dimensional theory
consisting of the Nambu-Goldstone boson and an infinite tower of
massive vector mesons. Their coupling constants are determined by only
two parameters of the original theory; the 't\,Hooft coupling
$\lambda$ and the mass scale $\MKK$ which specifies the radius of the
$S^1$ compactification of the D4-brane.

In the low energy limit of discarding all the massive vector mesons in
the SS-model, we get the Skyrme model \cite{Skyrme}, a low energy
effective theory of Nambu-Goldstone bosons.
The Skyrme model has a novel property that it has a stable soliton
solution describing baryons. Upon quantization of the collective
coordinate of the $SU(N_f)$ rotation of the baryon solution, the
static properties of nucleons such as masses, charge radii and the
magnetic moments have been computed \cite{ANW}. The results agree
fairly well with experiments with of course a number of exceptions.

The SS-model, namely, the five-dimensional YM+CS theory mentioned
above, also has a soliton solution which describes the baryon
\cite{HSSY,HRYY1,HRYY2}. This baryon solution at a time slice is
approximately a BPST instanton with a fixed size in flat
four-dimensional YM theory. In \cite{HSSY}, the collective coordinate
quantization of the baryon solution is carried out in the $N_f=2$ case
to give the baryon spectra containing negative-parity baryons as well
as the baryons with higher spins and isospins. By taking $\MKK$ of
about $500\MeV$, the obtained baryon spectra is in qualitatively good
agreement with experiment. The extension to the $N_f=3$ case is not so
straightforward: the original CS term of \cite{SaSu1,SaSu2} cannot
give the desired first class constraint which selects baryon states
with correct spins. In \cite{HMSU3}, the collective coordinate
quantization in the $N_f=3$ case was carried out by using a new CS
term which reproduces the desired constraint.

The purpose of this paper is to continue and accomplish the study of
static properties of nucleons in the SS-model with two
flavors.\footnote{
See \cite{HRYY2} and \cite{HRYY3}, for a different approach to the
properties of nucleons in holographic QCD.
}
We will examine the charge distributions, charge radii,
magnetic moments and the axial vector coupling, namely, the quantities
studied in \cite{ANW} in the $N_f=2$ Skyrme model.
The most nontrivial and difficult point in this study is how to
define, in the SS-model which is a five-dimensional YM+CS theory,
the chiral $U(N_f)_L\times U(N_f)_R$ current observed in our four
dimensional spacetime.
One way to define the chiral current is via the bulk-boundary
correspondence \cite{GKP,AdSCFTWitten}; we introduce the external
gauge fields on the boundaries $z=\pm\infty$ of the extra fifth
dimension $z$, and read off the current from the coupling with the
external fields.
The current obtained this way is defined in terms of fields on the
boundary. Therefore, it is invariant under gauge transformations which
does not change the boundary behavior of the fields.
However, we do not adopt this current in our study of static
properties of nucleons since the baryon solution is localized near the
origin $z=0$ of the extra fifth dimension and hence the current
vanishes for the baryon configuration (see, however, sec.\
\ref{subsec:compwtj} for a possibility of getting non-vanishing
results for static properties of nucleons).

In this paper, we take another definition of the chiral current.
For introducing it, let us recall that the chiral transformation in
four dimensions corresponding to
$(g_L,g_R)\in U(N_f)_L\times U(N_f)_R$
is realized in the SS-model in five dimensions as the local gauge
transformation taking the values $g_L$ and $g_R$ on the boundaries
$z=\infty$ and $-\infty$, respectively.
Therefore, we first consider in the SS-model the conserved Noether
current of the infinitesimal local gauge transformation which is
non-vanishing at $z=\pm\infty$. Then, the chiral current in four
dimensions is obtained by integrating the five-dimensional current
over $z$. However, the chiral current defined this way has two
problems which are related each other. First, this chiral current is
not determined uniquely by the gauge transformation on the boundary
$z=\pm\infty$, but it depends on the way of interpolating the local
gauge transformation for finite $z$. Second, this chiral current is
not a gauge invariant quantity. It is not invariant even under the
local gauge transformation that does not change the behavior of the
fields at $z=\pm\infty$.

Despite these problems, we adopt the chiral current defined from the
Noether current of local gauge transformation to compute the static
properties of nucleons. As the interpolating function of $z$ of the
gauge transformation, we take the zero-modes of the laplace operator
of $z$. There are two facts that supports our choice of chiral
current.
One is that it reproduces the chiral current of the Skyrme model in
the low energy limit of discarding all the massive $z$-modes.
The second fact is our result of the computation of the static
properties of nucleons itself. We find that our results in the
SS-model are generally closer to the experimental values than in the
Skyrme model \cite{ANW}. In particular, we get a surprising
improvement for the axial vector coupling compared with the Skyrme
model.

The organization of the rest of this paper is as follows.
In sec.\ \ref{sec:defj}, we introduce our chiral current from the
Noether current of local gauge transformation. We also discuss the
relationship of our current to the chiral current from the
bulk-boundary correspondence, and the problems of our current.
In sec.\ \ref{sec:SkyrmeLimit}, we consider the low energy limit of
our chiral current and show that it reproduces the chiral current of
the Skyrme model.
Sec.\ \ref{sec:staticprop} is the main part of this paper. We first
summarize the baryon classical solution in the SS-model and its
collective coordinate quantization. Then, we compute the various
static properties of nucleon using our chiral current.
In sec.\ \ref{sec:summary}, we summarize the paper and discuss
remaining problems. In appendix \ref{app:regulargauge}, we examine the
vector current in terms of the baryon solution in the regular gauge
to explain why we have to take the singular gauge solution in the
analysis of sec.\ \ref{sec:staticprop}.

\section{Four-dimensional chiral current in the SS-model}
\label{sec:defj}

For analyzing the static properties of nucleons such as the
electric charge density, magnetic moments and axial-vector coupling,
we have to first of all define the conserved
{\em four}-dimensional current of chiral $U(N_f)_L\times U(N_f)_R$
symmetry in the SS-model which is a YM+CS theory
in {\em five}-dimensions.
In this section, we propose a definition of the conserved chiral
current which will be used in later sections for the various analysis.
Our chiral current is given in terms of five-dimensional Noether
current of local gauge symmetry transformation which is non-vanishing
on the boundaries $z=\pm\infty$.
We will also discuss the relationship of our chiral current with
that defined via the bulk-boundary correspondence
\cite{GKP,AdSCFTWitten}, and the problems with our current, i.e.,
the uniqueness and the gauge-invariance.

\subsection{Action of the SS-model}

First, let us summarize the action of the SS-model, which is the
effective theory of $N_f$ probe D8-branes in the background of $N_c$
D4-branes. Concretely, it is a Yang-Mills (YM) theory in five
dimensions with gauge group $U(N_f)$ supplemented with the
Chern-Simons (CS) term:
\begin{align}
S[\cA] &= S_\YM[\cA]+S_\CS[\cA] ,
\label{eq:S=SYM+SCS}
\\
S_\YM[\cA]
&= -\kappa\int d^4xdz\tr\left[
\frac12 h(z)\cF_{\mu\nu}\cF^{\mu\nu}
+k(z)\cF_{\mu z}\cF^{\mu z}\right] ,
\label{eq:SYM}
\\
S_\CS[\cA]
&= \frac{N_c}{24\pi^2}\int\omega_5(\cA) ,
\label{eq:SCS}
\end{align}
where $\mu,\nu = 0,1,2,3$ are the four-dimensional Lorentz indices and
$z$ is the extra fifth dimension, and they are raised/lowered
by the flat metric $\eta_{MN}=\eta^{MN}=\diag(-1,1,1,1,1)$
($M,N=0,1,2,3,z$).
The one-form $\cA=\cA_\mu dx^\mu+\cA_z dz$ is the $U(N_f)$ gauge
field, and the corresponding field strength is given by
$\cF=d\cA+i\cA^2$. In $S_\YM$ \eqref{eq:SYM}, $\kappa$ is a constant
given in terms of the 't\,Hooft coupling $\lambda$ and the number
of colors $N_c$ as
\begin{equation}
\kappa = a\lambda N_c ,\quad
\left(a=\frac{1}{216\pi^3}\right),
\label{eq:kappa}
\end{equation}
and the warp factors $h(z)$ and $k(z)$ are
\begin{equation}
h(z) = \left(1+z^2\right)^{-1/3},\quad
k(z) = 1+z^2 .
\label{eq:h(z)k(z)}
\end{equation}
In the CS term $S_\CS$ \eqref{eq:SCS}, $\omega_5(\cA)$ is the CS
five-form:
\begin{equation}
\omega_5(\cA) =
\tr\left(\cA\cF^2-\frac{i}2\cA^3\cF-\frac1{10}\cA^5\right) .
\end{equation}
In \cite{HMSU3}, another expression of the CS term is proposed
by extending the gauge field $\cA$ to a six-dimensional space $M_6$
whose boundary $\p M_6$ is the original five-dimensional space
of the SS-model:
\begin{equation}
\SCSnew=\frac{N_c}{24\pi^2}\int_{M_6}\!\tr\cF^3 .
\label{eq:SCSnew}
\end{equation}
Although the two expressions \eqref{eq:SCS} and \eqref{eq:SCSnew} are
naively the same due to $\tr\cF^3=d\omega_5(\cA)$, it is crucial to
adopt \eqref{eq:SCSnew} in the case of $N_f=3$ for reproducing the
baryon states with correct spins \cite{HMSU3}. However, for the chiral
current in the $N_f=2$ case, there is essentially no difference
between the two CS terms.

The equations of motion (EOM) of $\cA_\mu$ and $\cA_z$
obtained from the action \eqref{eq:S=SYM+SCS} read respectively as
follows:
\begin{align}
2\kappa\left[
\cD_\nu\bigl(h(z)\cF^{\nu\mu}\bigr)
+\cD_z\bigl(k(z)\cF^{z\mu}\bigr)\right]
+\frac{N_c}{32\pi^2}\epsilon^{\mu NPQR}
\cF_{NP}\cF_{QR}&=0 ,
\label{eq:EOMA_mu}
\\
2\kappa\,\cD_\mu\bigl(k(z)\cF^{\mu z}\bigr)
+\frac{N_c}{32\pi^2}\epsilon^{\mu\nu\rho\sigma}
\cF_{\mu\nu}\cF_{\rho\sigma}&=0 ,
\label{eq:EOMA_z}
\end{align}
with $\epsilon^{0123z}=\epsilon^{0123}=1$.
Here, we have used that the CS five-form changes under an
infinitesimal deformation $\delta\cA$ of the gauge field $\cA$ as
\begin{equation}
\delta\omega_5(\cA)=3\tr\bigl(\delta\cA\,\cF^2\bigr)
+d\beta\bigl(\delta\cA,\cA\bigr) ,
\label{eq:deltaomega_5}
\end{equation}
with $\beta$ given by
\begin{equation}
\beta\bigl(\delta\cA,\cA\bigr)=\tr\left[\delta\cA
\left(\cF\cA+\cA\cF-\frac{i}{2}\cA^3\right)\right] .
\end{equation}
The last $\epsilon$-tensor terms in \eqref{eq:EOMA_mu} and
\eqref{eq:EOMA_z} remain unchanged even if we adopt another CS term
\eqref{eq:SCSnew}.

\subsection{Conserved current of local gauge symmetry}
\label{subsec:constj}

In order to propose a definition of the chiral current in the
SS-model, let us first recall how the $U(N_f)_L\times U(N_f)_R$
symmetry in four dimensions is realized in the model. In the SS-model
in a gauge with the boundary condition,
\begin{equation}
\cA_M(x,z)\to 0 \quad (z\to\pm\infty) ,
\label{eq:cA_M->0}
\end{equation}
the pion field $U(x)$ in four dimensions is defined by
\cite{SaSu1,SaSu2}
\begin{equation}
U(x) =
\Po\exp\left(-i\int^\infty_{-\infty}\!dz\,\cA_z(x,z)\right) ,
\label{eq:U(x)}
\end{equation}
where the path-ordering is from the right at $z=-\infty$ to the left
at $z=+\infty$.
Then, the chiral transformation corresponding to
$(g_L,g_R)\in U(N_f)\times U(N_f)_R$ is realized as the
local gauge transformation
\begin{equation}
\cA \to \cA^g = g(\cA-id)g^{-1} ,
\label{eq:A^g}
\end{equation}
with $g(x,z)\in U(N_f)$ which tends to constants as $z\to\pm\infty$,
\begin{equation}
g(x,z)\to
\begin{cases}
g_L & (z\to +\infty)\\ g_R & (z\to -\infty)
\end{cases} ,
\end{equation}
and hence keeps the boundary condition \eqref{eq:cA_M->0}.
In fact, under this local gauge transformation, the pion field
transforms as
\begin{equation}
U(x) \to g_LU(x)g_R^{-1} .
\label{eq:U->gLUgR^-1}
\end{equation}

Motivated by the realization of the chiral symmetry transformation in
the SS-model as a local gauge transformation, we propose the following
construction of the conserved chiral current in four dimensions:
\begin{enumerate}
\item
First, let us consider the conserved Noether current in
{\em five}-dimensions, $\fJ_\gtfn^M(x,z)$, corresponding to the
infinitesimal local gauge symmetry transformation,
\begin{equation}
\delta_\gtfn\cA_M = \cD_M\gtfn = \p_M\gtfn+i[\cA_M,\gtfn] ,
\end{equation}
with $\gtfn(x,z)$ satisfying the boundary condition
\begin{equation}
\gtfn(x,z)\to
\begin{cases}
\gtfn_L=\gtfn_L^a\,t_a & (z\to +\infty) \\
\gtfn_R=\gtfn_R^a\,t_a & (z\to -\infty)
\end{cases} ,
\label{eq:BCzeta(x,z)}
\end{equation}
where $\gtfn_{L/R}^a$ are arbitrary constants, and $t_a$
($a=0,1,2,\cdots,N_f^2-1$) is the hermitian generator of
$U(N_f)$ satisfying
\begin{equation}
\left[t_a,t_b\right]=i f_{abc}t_c ,\quad
\tr\bigl(t_a t_b\bigr)=\frac12\,\delta_{ab} ,\quad
t_0=\frac{1}{\sqrt{2 N_f}}\,\bm{1}_{N_f} .
\end{equation}

\item
Then, we define the conserved {\em four}-dimensional current
$j_\gtfn^\mu(x,z)$ by
\begin{equation}
j_\gtfn^\mu(x)=\int_{-\infty}^\infty\! dz\,\fJ_\gtfn^\mu(x,z) .
\label{eq:j^zeta_mu}
\end{equation}
Since $\fJ_\gtfn^M(x,z)$ is conserved in the five-dimensional sense,
\begin{equation}
\p_M\fJ_\gtfn^M=\p_\mu\fJ_\gtfn^\mu+\p_z\fJ_\gtfn^z=0 ,
\label{eq:p_McJ^zeta_M=0}
\end{equation}
the four-dimensional current $j_\gtfn^\mu(x)$ satisfies the
conservation law
\begin{equation}
\p_\mu j_\gtfn^\mu=0 ,
\end{equation}
provided the $z$-component $\fJ_\gtfn^z$ vanishes on the boundary:
\begin{equation}
\fJ_\gtfn^z(x,z\to\pm\infty)=0 .
\label{eq:cJ_z->0}
\end{equation}

\item
It follows (using \eqref{eq:deltaA} and \eqref{eq:deltaS} given below)
that the five-dimensional current $\fJ^\gtfn_M(x,z)$ and the pion
field $U(x)$ \eqref{eq:U(x)} satisfy the Ward identity\footnote{
If we take into account the gauge-fixing and the Faddeev-Popov ghost
terms necessary for the quantization, the current $\fJ^\gtfn_M$ is no
longer conserved for a generic $\gtfn$ but its divergence is given by
a BRST-exact form \cite{KugoOjima,RLGS}:
$$
\p_M\fJ_\gtfn^M=\left\{\QB,*\right\} ,
$$
with $\QB$ being the BRST charge.
However, the Ward identities \eqref{eq:WIfive} and \eqref{eq:WIfour}
remain valid since the pion field is a physical operator satisfying
$\left[\QB,U(x)\right]=0$ (we assume that the Faddeev-Popov ghost
$c(x,z)$ vanishes at infinity).
}
\begin{equation}
\p_M\bVEV{\fJ_\gtfn^M(x,z)\,U(x')\,\cO}
=-\delta^4(x-x')\bVEV{
U(x;\infty,z)\cD_z\gtfn(x,z)\,U(x;z,-\infty)\cO}
+\ldots\ ,
\label{eq:WIfive}
\end{equation}
where $U(x;z_1,z_2)$ is defined by
\begin{equation}
U(x;z_1,z_2)=
\Po\exp\left(-i\int^{z_1}_{z_2}\!dz\,\cA_z(x,z)\right) ,
\end{equation}
and $\cO$ and the dots on the RHS represent other operators and the
terms containing their gauge transformation, respectively.
Integrating \eqref{eq:WIfive} over $z$ and using \eqref{eq:cJ_z->0},
we obtain
\begin{equation}
\p_\mu\VEV{j_\gtfn^\mu(x)\,U(x')\,\cO}
=-\delta^4(x-x')\VEV{\bigl(\gtfn_L U(x)-U(x)\gtfn_R\bigr)
\,\cO}+\ldots\ .
\label{eq:WIfour}
\end{equation}
This is the desired chiral Ward identity in four dimensions. The
four-dimensional current $j^\gtfn_\mu(x)$ with $\gtfn_{R/L}=0$ is the
left/right-current.

\end{enumerate}

The conserved five-dimensional current $\fJ^\gtfn_M(x,z)$ is obtained
by making on the action the infinitesimal transformation,
\begin{equation}
\delta\cA_M(x,z) = \veps(x,z)\cD_M\gtfn(x,z) ,
\label{eq:deltaA}
\end{equation}
with $\veps(x,z)$ vanishing at infinity, and using the
identification,
\begin{equation}
\delta S=\int\!d^4x dz \,\fJ_\gtfn^M(x,z)\,\p_M\veps(x,z) .
\label{eq:deltaS}
\end{equation}
For the action \eqref{eq:S=SYM+SCS}, we get
\begin{align}
\fJ_\gtfn^M = \fJ_{\YM\,\gtfn}^M+\fJ_{\CS\,\gtfn}^M ,
\label{eq:cJ^zeta_M}
\end{align}
with $\fJ_{\YM\,\gtfn}^M$ and $\fJ_{\CS\,\gtfn}^M$ from $S_\YM$
\eqref{eq:SYM} and $S_\CS$ \eqref{eq:SCS}, respectively, given by
\begin{align}
\fJ_{\YM\,\gtfn}^\mu(x,z)
&=-2\kappa\tr\bigl(
h(z)\cF^{\mu\nu}\cD_\nu\gtfn+k(z)\cF^{\mu z}\cD_z\gtfn\bigr) ,
\label{eq:cj^zeta_muYM}
\\
\fJ_{\YM\,\gtfn}^z(x,z)
&=-2\kappa\,k(z)\tr\bigl(\cF^{z\nu}\cD_\nu\gtfn\bigr) ,
\\
\fJ_{\CS\,\gtfn}^M(x,z)&=-\frac{N_c}{64\pi^2}\,
\veps^{MNPQR}\tr\bigl(\{\cF_{NP},\cF_{QR}\}\gtfn\bigr) .
\label{eq:cJ^zeta_MCS}
\end{align}
Using the EOM, \eqref{eq:EOMA_mu} and \eqref{eq:EOMA_z}, we can
confirm that $\fJ_\gtfn^M$ satisfies the conservation equation
\eqref{eq:p_McJ^zeta_M=0}.

\subsection{Relation to the current from bulk-boundary correspondence}
\label{subsec:OtherJ}

In this subsection, let us consider the relationship of our chiral
current introduced above with the chiral current
$\wt{j}^{L/R}_\mu(x)$ defined a la the bulk-boundary correspondence
\cite{GKP,AdSCFTWitten}. The latter is obtained as follows. We solve
the EOM under the boundary condition,
\begin{equation}
\cA_\mu(x,z)\to
\begin{cases}
\cA^L_{\mu}(x) & (z\to +\infty)\\
\cA^R_{\mu}(x) & (z\to -\infty)
\end{cases} ,
\label{eq:A->A^L/R}
\end{equation}
insert the solution depending on $\cA^{L/R}_\mu$ into the action,
and read off the currents $\wt{j}_{L/R}^\mu(x)$ from the coupling
\begin{equation}
\int\!d^4x\,\tr\left(\cA^L_{\mu}(x)\,\wt{j}_L^{\mu}(x)
+\cA^R_{\mu}(x)\,\wt{j}_R^{\mu}(x)\right) .
\label{eq:coupling}
\end{equation}
This is equivalent to making the following shift in the action,
\begin{equation}
\cA_\mu(x,z)\to
\cA_\mu(x,z) +\cA^L_\mu(x)\psi_L(z)+\cA^R_\mu(x)\psi_R(z) ,
\label{eq:shiftcA_SS}
\end{equation}
with $\psi_{L/R}(z)$ being functions satisfying the boundary
condition
\begin{equation}
\psi_{L/R}(z)\to
\begin{cases}
  1 & (z\to\pm\infty) \\ 0 & (z\to\mp\infty)
\end{cases}
,
\end{equation}
and keeping terms linear in $\cA^{L/R}_\mu(x)$ by using that
$\cA_M(x,z)$ is a solution to the EOM with the boundary condition
\eqref{eq:cA_M->0}.
We find that the chiral current $\wt{j}_{L/R}^\mu(x)$ obtained
this way is\footnote{
In addition to \eqref{eq:wtj} from $S_\YM$ \eqref{eq:SYM}, we have
another contribution
$$
\wt{j}_{\CS\,L/R}^{\mu}(x)
=\mp\frac{N_c}{48\pi^2}\,\epsilon^{\mu\nu\rho\sigma}
\Bigl(\cF_{\nu\rho}\cA_\sigma
+\cA_\nu\cF_{\rho\sigma}-i\cA_\nu\cA_\rho\cA_\sigma\Bigr)
\Bigr|_{z=\pm\infty} ,
$$
from the CS term \eqref{eq:SCS} due to the $d\beta$ term in
\eqref{eq:deltaomega_5}. Here, we have ignored this
$\wt{j}_{\CS\,L/R}^{\mu}(x)$ since it vanishes at $z=\pm\infty$,
in contrast with \eqref{eq:wtj} which is multiplied by $k(z)$
\eqref{eq:h(z)k(z)}. We have no $\wt{j}_{\CS\,L/R}^{\mu}(x)$
at all in the case of another CS term \eqref{eq:SCSnew} since we
simply have $\delta\tr\cF^3=3\,d\tr\bigl(\delta\cA\,\cF^2\bigr)$.
}
\begin{equation}
\wt{j}_{L/R}^\mu(x)=\mp 2\kappa\,k(z)\cF^{\mu z}(x,z)
\Bigr|_{z=\pm\infty} .
\label{eq:wtj}
\end{equation}
The relation between the two currents, \eqref{eq:j^zeta_mu} and
\eqref{eq:wtj}, is as follows. First, the five-dimensional current
\eqref{eq:cJ^zeta_M} is rewritten as
\begin{align}
\fJ_\gtfn^{\mu}(x,z)
&=-2\kappa\,\p_\nu\tr\bigl[h(z)\cF^{\mu\nu}(x,z)\gtfn(x,z)\bigr]
\nn\\
&\quad
-2\kappa\,\p_z\tr\bigl[k(z)\cF^{\mu z}(x,z)\gtfn(x,z)\bigr]
-\tr\bigl[(\mbox{LHS of \eqref{eq:EOMA_mu}})\gtfn(x,z)\bigr] .
\label{eq:cJ=p(...)}
\end{align}
Integrating this over $z$, we find that
\begin{equation}
j_\gtfn^\mu(x)=\tr\!\left(\gtfn_L\,\wt{j}_L^{\mu}(x)
+\gtfn_R\,\wt{j}_R^{\mu}(x)\right)
+\p_\nu\chi^{\mu\nu}(x)+(\mbox{EOM-term}) ,
\label{eq:relation}
\end{equation}
where $\gtfn_{L/R}$ are the Lie-algebra valued constants of
\eqref{eq:BCzeta(x,z)}, and the anti-symmetric tensor
$\chi^{\mu\nu}(x)$ is defined by
\begin{equation}
\chi^{\mu\nu}(x) = -\chi^{\nu\mu}(x)
=-2\kappa\int^\infty_{-\infty}\!dz
\tr\bigl[h(z)\cF^{\mu\nu}(x,z)\gtfn(x,z)\bigr] .
\label{eq:chi^munu}
\end{equation}
This relation implies that the two currents, $j_\gtfn^\mu(x)$ and
$\wt{j}_{L/R}^{\mu}(x)$, are ``equivalent'' in the sense that their
difference is an identically conserving term $\p_\nu\chi^{\mu\nu}$.
In particular, the integrated charges $Q=\int\! d^3x\, j^0(x)$ are the
same between the two if we are allowed to discard the surface integral
from the $\p_i\chi^{0i}$ term. However, the local currents themselves
do differ from each other, and this difference can lead to different
physics.
Moreover, we will see in sec.\ \ref{sec:staticprop} that, in the
quantization of baryons, $\wt{j}_{L/R}^\mu(x)$ vanishes and the
main contribution to $j_\gtfn^\mu(x)$ \eqref{eq:relation} is from the
$\p_\nu\chi^{\mu\nu}(x)$ term.

\subsection{Gauge non-invariance of $\bm{j^\gtfn_\mu}$
and non-uniqueness of $\bm{\gtfn}$}
\label{subsec:problemsofj}

Here, we discuss the problems inherent in our four-dimensional chiral
current $j_\gtfn^\mu(x)$ \eqref{eq:j^zeta_mu} itself. First recall
that our chiral current should be considered, by its construction, for
the gauge fields $\cA_M$ satisfying the boundary condition
\eqref{eq:cA_M->0}.
The most serious problem with our chiral current $j_\gtfn^\mu(x)$ is
that it is not a gauge invariant quantity. It is not invariant even
under the gauge transformation which keeps the boundary condition
\eqref{eq:cA_M->0}. The integrated conserved charge
$Q_\gtfn=\int\!d^3x\,j_\gtfn^0(x)$, which can be expressed as a
spatial surface integration as seen from \eqref{eq:cJ=p(...)}, is
invariant under the gauge transformation which does not change the
boundary behavior of $\cA$. However, since we have to treat the local
currents $j_\gtfn^\mu(x)$ itself in the analysis of static properties
of nucleons, the gauge-noninvariance of our chiral current is an
important problem to be resolved.

Another and related problem with our chiral current is the
non-uniqueness of the function $\gtfn(x,z)$ defining the current.
In the construction of our chiral current given in sec.\
\ref{subsec:constj}, the only condition on $\gtfn(x,z)$ is the
boundary condition \eqref{eq:BCzeta(x,z)}, which of course cannot
uniquely determine $\gtfn(x,z)$.
This problem of the non-uniqueness of $\gtfn(x,z)$ is intimately
related with the gauge-noninvariance of our chiral current.
Note that the five-dimensional current $\fJ_\gtfn^M$
\eqref{eq:cJ^zeta_M} and hence the four-dimensional one
$j_\gtfn^\mu(x)$ are invariant under the simultaneous gauge
transformation of both the gauge field $\cA_M(x,z)$ and $\gtfn(x,z)$:
\begin{align}
\cA_M&\to \cA^g=g\bigl(\cA_M-i\p_M\bigr)g^{-1} ,
\nn\\
\gtfn&\to \gtfn^g=g\,\gtfn g^{-1} .
\label{eq:gtfn^g}
\end{align}
This fact implies that we cannot choose a specific function
$\gtfn(x,z)$ for our chiral current: For giving the same chiral
current among different $\cA$'s related to each other via gauge
transformations which keep the boundary behavior, we have to adjust
the corresponding $\gtfn(x,z)$ according to the formula
\eqref{eq:gtfn^g}.

On the other hand, the current $\wt{j}_{L/R}^\mu(x)$ \eqref{eq:wtj}
of the bulk-boundary correspondence are free from both of the above
two problems. In particular, since it is defined by the fields on
the boundaries $z=\pm\infty$, it is inert under local gauge
transformations which do not change the boundary behavior of the
fields. However, the current $\wt{j}_{L/R}^\mu(x)$ does not seem to
give physically sensible results. First, $\wt{j}_{L/R}^\mu(x)$ can
reproduce only a part of the chiral current of the Skyrme model in
the low energy limit (see the end of sec.\ \ref{eq:LELj}).
Second, $\wt{j}_{L/R}^\mu(x)$ vanishes for the baryon configuration
since the baryon is localized at the origin $z=0$ and cannot be seen
on the boundary (see the end of sec.\ \ref{subsec:quantization}).

We do not have definite solutions to the problems of our chiral
current $j_\gtfn^\mu(x)$. However, in the following sections,
we continue our analysis by using the present chiral current
\eqref{eq:j^zeta_mu} with $\fJ_\gtfn^M$ given by
\eqref{eq:cJ^zeta_M}--\eqref{eq:cJ^zeta_MCS}.
For definiteness, we adopt as $\gtfn(x,z)$ the following one:
\begin{equation}
\gtfn(x,z)=\psi_\pm(z) t_a ,
\label{eq:ourzeta(x,z)}
\end{equation}
where $\psi_\pm(z)$ are the zero-modes of the differential operator
$-k(z)^{1/3}\p_z k(z)\p_z$ \cite{SaSu1,SaSu2}:
\begin{equation}
\psi_\pm(z)=\frac12\pm\frac{1}{\pi}\arctan z
\to
\begin{cases}
1 & (z\to\pm\infty)\\ 0 & (z\to\mp\infty)
\end{cases}.
\label{eq:psi_pm}
\end{equation}
This $\psi_\pm(z)$ has been adopted in \cite{SaSu1,SaSu2} as
$\psi_{L/R}(z)$ of \eqref{eq:shiftcA_SS} in introducing the external
fields $\cA^{L/R}_\mu(x)$.\footnote{
If we take $\cA^{L/R}_\mu$ of the form
$\cA^{L/R}_\mu(x)=\p_M\veps^{L/R}_a(x)\,t_a$, the shift
\eqref{eq:shiftcA_SS} is essentially equivalent to the infinitesimal
transformation \eqref{eq:deltaA} with $\gtfn(x,z)$ given by
\eqref{eq:ourzeta(x,z)} with the identification
$\psi_{L/R}(z)=\psi_\pm(z)$:
The shift of $\cF_{MN}$ under \eqref{eq:shiftcA_SS} is
$\p_{[N}\veps^L_a(x)\cD_{M]}\bigl(\psi_+(z)t_a\bigr)
+\p_{[N}\veps^R_a(x)\cD_{M]}\bigl(\psi_-(z)t_a\bigr)$,
while we have
$\delta\cF_{MN}=\veps\left[\cF_{MN},\gtfn\right]
+(\p_{[M}\veps)\cD_{N]}\gtfn$ under \eqref{eq:deltaA}.
}
Since $\psi_\pm(z)$ is the zero-mode, no mass terms are developed for
$\cA^{L/R}_\mu(x)$.
However, we have no convincing reason for adopting $\psi_\pm(z)$
in the context of the Noether current of local gauge transformation.
The only reasons why we adopt our chiral current \eqref{eq:j^zeta_mu}
with $\gtfn(x,z)$ of \eqref{eq:ourzeta(x,z)} are that it reproduces
in the low energy limit the chiral current of the Skyrme model,
and that the various static properties of nucleons are obtained as
non-vanishing and finite numbers which are close to the experimental
values, as we will see in secs.\ \ref{sec:SkyrmeLimit} and
\ref{sec:staticprop}, respectively.

Explicitly, the chiral current $j_{L/R,a}^\mu(x)$ corresponding to
\eqref{eq:ourzeta(x,z)} is given by
\begin{equation}
j_{L/R,a}^\mu(x)=\int_{-\infty}^\infty\!dz\,\fJ_{L/R,a}^\mu(x,z) ,
\label{eq:ourj=intourJ}
\end{equation}
with
\begin{align}
\fJ_{L/R,a}^\mu(x,z)&=-2i\kappa\tr\Bigl\{\Bigl(
h(z)\left[\cF^{\mu\nu},\cA_\nu\right]
+k(z)\left[\cF^{\mu z},\cA_z\right]\Bigr)t_a\Bigr\}\psi_\pm(z)
\nn\\
&\quad
-2\kappa\,k(z)\tr\bigl(\cF^{\mu z}t_a\bigr)\drv{\psi_\pm(z)}{z}
-\frac{N_c}{64\pi^2}\,
\veps^{\mu NPQR}\tr\bigl(\{\cF_{NP},\cF_{QR}\}t_a\bigr)\psi_\pm(z) .
\label{eq:ourJ_mu}
\end{align}
The vector and the axial-vector currents are obtained by replacing
$\psi_\pm(z)$ in \eqref{eq:ourJ_mu} with the following $\psi_V(z)$ and
$\psi_A(z)$, respectively:
\begin{equation}
\psi_V(z)=\psi_+(z)+\psi_-(z)=1,\quad
\psi_A(z)=\psi_+(z)-\psi_-(z)=\frac{2}{\pi}\arctan z .
\label{eq:psi_VA}
\end{equation}
In this case, the $z$-component of the five-dimensional
current which we have to confirm to vanish at $z=\pm\infty$ (recall
\eqref{eq:cJ_z->0}) is
\begin{equation}
J_{L/R,a}^z(x,z)=-2i\kappa\,k(z)\tr\bigl([\cF_{z\nu},\cA_\nu]t_a
\bigr)\psi_\pm(z)
-\frac{N_c}{32\pi^2}\,\epsilon^{\mu\nu\rho\sigma}
\tr\bigl(\cF_{\mu\nu}\cF_{\rho\sigma}t_a\bigr)\psi_\pm(z) .
\label{eq:ourJ_z}
\end{equation}

\section{Chiral currents in the Skyrme approximation}
\label{sec:SkyrmeLimit}

In the last section, we proposed a definition of the four-dimensional
chiral current $j_\gtfn^\mu(x)$ \eqref{eq:j^zeta_mu}. It is given as
the $z$-integration of the five-dimensional Noether current
$\fJ_\gtfn^\mu(x,z)$ of the local gauge symmetry transformation with
the boundary condition \eqref{eq:BCzeta(x,z)}.
In this section, as a test of the validity of this definition of the
chiral current in the SS-model, we evaluate it in the Skyrme
approximation (i.e., in the low energy limit). It has been
known that the SS-model is reduced to the Skyrme model in the low
energy limit \cite{SaSu1,SaSu2}.
The Skyrme model is an ordinary four-dimensional field theory
having the pion field (Skyrme field) $U(x)\in U(N_f)$ as its dynamical
variable, and the chiral currents of the symmetry transformation
\eqref{eq:U->gLUgR^-1} is simply the corresponding Noether current.
Using this Noether current of chiral symmetry, various properties of
the model including the static properties of nucleons have been
analyzed \cite{ANW}.
What we wish to test here is whether the chiral current
proposed in sec.\ \ref{sec:defj} is reduced to the chiral current in
the Skyrme model.

\subsection{SS-model in the low energy limit}

Before considering the low energy limit of our chiral current in the
SS-model, we in this subsection review the derivation of the Skyrme
model as the low energy limit of the SS-model.
For this purpose, it is convenient to move, from the gauge with the
boundary condition \eqref{eq:cA_M->0}, to the gauge with
$\cA_z(x,z)=0$.
Let us take as the function $g(x,z)\in U(N_f)$ in \eqref{eq:A^g} for
realizing $\cA^g_z(x,z)=0$ the following one:
\begin{equation}
g(x,z)^{-1}=\Po\exp\left(-i\int_{-\infty}^z\! dz'\cA_z(x,z')\right) .
\label{eq:g(x,z)^-1}
\end{equation}
In the rest of this section, the gauge fields $\cA$ and $\cA^g$ denote
the one with the boundary condition \eqref{eq:cA_M->0} and that in the
$\cA^g_z=0$ gauge, respectively.

In the low energy limit of the SS-model in the $\cA^g_z=0$ gauge, we
mode-expand the $z$-dependence of $\cA^g_\mu(x,z)$ in terms of the
eigenfunctions of $-k(z)^{1/3}\p_z k(z)\p_z$ and keep only the
zero-modes $\psi_\pm(z)$ \eqref{eq:psi_pm}.
Taking into account that
\begin{equation}
\cA^g_\mu(x,z)\to
\begin{cases}
R_\mu(x) & (z\to +\infty)\\ 0 & (z\to -\infty)
\end{cases} ,
\end{equation}
with $R_\mu(x)$ defined by
\begin{equation}
R_\mu(x)=-iU(x)^{-1}\p_\mu U(x) ,
\end{equation}
we have in the present approximation
\begin{equation}
\cA^g_\mu(x,z)=R_\mu(x)\psi_+(z) ,
\label{eq:A^g_mu=}
\end{equation}
where all the massive-modes have been dropped on the RHS.
Then, the field strengths are given by
\begin{align}
\cF^g_{\mu\nu}(x,z)&=-i\left[R_\mu(x),R_\nu(x)\right]
\psi_+(z)\psi_-(z) ,
\label{eq:F^g_munu=}
\\
\cF^g_{z\nu}(x,z) &= R_\nu(x)\drv{\psi_+(z)}{z} .
\label{eq:F^g_znu=}
\end{align}
Plugging the expressions \eqref{eq:F^g_munu=} and \eqref{eq:F^g_znu=}
into the YM part action \eqref{eq:SYM} with $\cF_{MN}$ replaced by
the gauge transformed one $\cF^g_{MN}$ and carrying out the
$z$-integration, we get the four-dimensional Skyrme lagrangian of the
field $U(x)$ \eqref{eq:U(x)} in the low energy limit:
\begin{equation}
\cL_U=\kappa\tr\!\left(-\frac{1}{\pi}R_\mu R^\mu
+\frac{c_S}{2}\,\left[R_\mu,R_\nu\right]
\left[R^\mu,R^\nu\right]\right) ,
\label{eq:cL_U}
\end{equation}
where $c_S$ is a constant given by
\begin{equation}
c_S=\int_{-\infty}^\infty\! dz\,h(z)
\left[\psi_+(z)\psi_-(z)\right]^2=0.156\ldots\ .
\label{eq:cS}
\end{equation}

The expression of the low energy limit of the CS term depends on
whether we adopt the original one \eqref{eq:SCS} or another one
\eqref{eq:SCSnew} proposed in \cite{HMSU3} (see sec.\ 5.5 of
\cite{SaSu1} for the former, and appendix D of \cite{HMSU3} for the
latter).
In both the cases, the CS term is reduced in the low energy limit to
$N_c\Gamma_{\rm WZW}[U]$ with the Wess-Zumino-Witten term
$\Gamma_{\rm WZW}[U]$ given by an apparently common form:
\begin{equation}
\Gamma_{\rm WZW}[U]
=\frac{1}{240\pi^2}\int\!\tr\bigl(-iU^{-1}dU\bigr)^5 .
\label{eq:G_WZW}
\end{equation}
For the original CS term of \eqref{eq:SCS}, the integration is over
the five-dimensional space-time of $(x^\mu,z)$, and
$U$ is $g(x,z)$ of \eqref{eq:g(x,z)^-1}. In the case of
\eqref{eq:SCSnew}, the integration of \eqref{eq:G_WZW} is over the
five-dimensional space-time, $M_6$ with the $z$ part removed, and $U$
is given by \eqref{eq:U(x)} with $\cA_z$ replaced by that on $M_6$.
In summary, the low energy limit of the SS-model is the Skyrme model
with the WZW term:
\begin{equation}
S_{\rm Skyrme}[U]=\int\!d^4x\,\cL_U + N_c\Gamma_{\rm WZW}[U] .
\label{eq:S_Skyrme}
\end{equation}

\subsection{Low energy limit of the chiral currents in the SS-model}
\label{eq:LELj}

Now we wish to show that our chiral current in the SS-model,
$j_{L/R,a}^\mu(x)$ \eqref{eq:ourj=intourJ} with \eqref{eq:ourJ_mu},
reduces in the low energy limit to the corresponding one in the Skyrme
model \eqref{eq:S_Skyrme}. For this purpose we first look more
carefully at the process of moving to the $\cA^g_z=0$ gauge.

In the low energy limit of the SS-model in a gauge with the boundary
condition \eqref{eq:cA_M->0}, $\cA_z(x,z)$ has only the mode
$\wh{\phi}_0(z)$,
\begin{equation}
\wh{\phi}_0(z)=\pm\drv{\psi_\pm(z)}{z}=\frac{1}{\pi}\frac{1}{1+z^2} ,
\label{eq:whphi_0}
\end{equation}
and is given by
\begin{equation}
\cA_z(x,z)=\varphi(x)\wh{\phi}_0(z),
\end{equation}
in terms of a Lie-algebra valued function $\varphi(x)$.
Then, $g(x,z)^{-1}$ \eqref{eq:g(x,z)^-1} for moving to the $\cA^g_z=0$
gauge and the pion field $U(x)$ \eqref{eq:U(x)} are given in terms of
$\varphi(x)$ as
\begin{align}
&g(x,z)^{-1}=\exp\bigl(-i\varphi(x)\psi_+(z)\bigr) ,
\label{eq:g^-1=exp}
\\
&U(x)=g(x,z=\infty)^{-1}=\exp\bigl(-i\varphi(x)\bigr) .
\end{align}
For this $g(x,z)^{-1}$ we actually have
\begin{equation}
\cA^g_z=g\left(\cA_z-i\p_z\right)g^{-1}=0 .
\label{eq:g(A_z-ip_z)g^-1=0}
\end{equation}
The key equations used in the rest of this subsection are the
projector-like properties of the zero-modes $\psi_\pm(z)$ in the low
energy approximation of dropping the massive $z$-modes:
\begin{equation}
\psi_\pm(z)^2\simeq \psi_\pm(z),\quad
\psi_+(z)\psi_-(z)\simeq 0,\quad
\psi_+(z)+\psi_-(z)=1 .
\label{eq:Projprop}
\end{equation}
The first equation is due to that a function $f(z)$ with the boundary
behavior $f(z\to\pm\infty)=1$ and $f(z\to\mp\infty)=0$ is
mode-expanded as $f(z)=\psi_\pm(z)+\mbox{massive-modes}$, while the
second one is due to that $f(z)$ which vanish both at $z=\pm\infty$
contains only the massive-modes. The third equation is an exact one.
Using the first equation, $g(x,z)^{-1}$ \eqref{eq:g^-1=exp} is
approximated as
\begin{equation}
g(x,z)^{-1}=1+\sum_{n=1}^\infty\frac{1}{n!}
\bigl(-i\varphi(x)\psi_+(z)\bigr)^n
\simeq
1+\sum_{n=1}^\infty\frac{1}{n!}\bigl(-i\varphi(x)\bigr)^n\psi_+(z)
=\psi_-(z)+U(x)\psi_+(z) .
\label{eq:Appg^-1}
\end{equation}
Likewise we have
\begin{equation}
g(x,z)\simeq \psi_-(z)+U(x)^{-1}\psi_+(z) ,
\label{eq:Appg}
\end{equation}
and we can confirm $g^{-1}g\simeq 1$ by using \eqref{eq:Projprop}.
Note, however, that it is dangerous to use \eqref{eq:Appg^-1} and
\eqref{eq:Appg} in expressions containing the derivatives with
respect to $z$, such as \eqref{eq:g(A_z-ip_z)g^-1=0}.
For example, differentiating $\psi_+^n\simeq\psi_+$ ($n\ge 2$) with
respect to $z$ and using again \eqref{eq:Projprop}, we get a
conflicting relation $n\psi_+\wh{\phi}_0\simeq \wh{\phi}_0$.

Having finished a preparation, let us turn to considering our chiral
current in the SS-model in the low energy limit.
Note first that our chiral current $j_\gtfn^\mu(x)$ with $\gtfn(x,z)$
of \eqref{eq:ourzeta(x,z)} has a meaning for configurations
satisfying the boundary condition \eqref{eq:cA_M->0}.
For using the low energy expressions \eqref{eq:A^g_mu=},
\eqref{eq:F^g_munu=} and \eqref{eq:F^g_znu=} in the $\cA^g_z=0$ gauge,
we use the fact that $j_\gtfn^\mu(x)$ is invariant under the
simultaneous gauge transformations \eqref{eq:gtfn^g} of both $\cA_M$
and $\gtfn$. For $\gtfn=\psi_-t_a$ corresponding to the right-current
$j_{R,a}^\mu(x)$, $\gtfn^g$ in the $\cA^g_z=0$ gauge is the same as
the original one in the low energy limit:
\begin{equation}
\gtfn^g\simeq\bigl(\psi_-+U^{-1}\psi_+\bigr)\psi_- t_a
\bigl(\psi_- +U\psi_+\bigr)
\simeq \psi_- t_a ,
\end{equation}
where we have used the expressions \eqref{eq:Appg^-1} and
\eqref{eq:Appg} for $g^{-1}$ and $g$, respectively, and the projector
property \eqref{eq:Projprop}.
Therefore, $j_{R,a}^\mu(x)$ in the low energy limit is obtained by
simply replacing $\cA_\mu$ and $\cF_{MN}$ in \eqref{eq:ourJ_mu} with
those in the $\cA^g_z=0$ gauge, \eqref{eq:A^g_mu=},
\eqref{eq:F^g_munu=} and \eqref{eq:F^g_znu=}, and carrying out the
integration over $z$:
\begin{equation}
j_{R,a}^\mu(x)\simeq
-2\kappa\tr\left\{\left(\frac{1}{\pi}R^\mu
+c_S\bigl[[R^\mu,R^\nu],R_\nu\bigr]\right)t_a\right\}
-\frac{i N_c}{48\pi^2}\epsilon^{\mu\nu\rho\lambda}\tr\bigl(
R_\nu R_\rho R_\lambda t_a\bigr) .
\label{eq:j^Ra_mu_SK}
\end{equation}
This agrees with the Noether current of the right-transformation
$U(x)\to U(x)g^{-1}_R$ in the Skyrme model \eqref{eq:S_Skyrme}.
It is obvious that the $z$-component of the five-dimensional current
\eqref{eq:ourJ_z} vanishes at $z=\pm\infty$ and hence satisfies the
condition \eqref{eq:cJ_z->0} necessary for the conservation of the
four-dimensional current \eqref{eq:j^Ra_mu_SK}.

Next, let us consider the left-current $j_{L,a}^\mu(x)$ corresponding
to $\gtfn=\psi_+ t_a$. In this case, the gauge transformation to the
$\cA^g_z=0$ gauge effects a nontrivial change on $\gtfn$:
\begin{equation}
\gtfn^g\simeq\bigl(\psi_-+U^{-1}\psi_+\bigr)\psi_+ t_a
\bigl(\psi_- +U\psi_+\bigr)
\simeq \psi_+(z) U(x)^{-1}t_a U(x) .
\end{equation}
Plugging this $\gtfn^g$ together with \eqref{eq:A^g_mu=},
\eqref{eq:F^g_munu=} and \eqref{eq:F^g_znu=} into
\eqref{eq:cj^zeta_muYM} and \eqref{eq:cJ^zeta_MCS}, and
using, in particular, that
$\cD^g_\nu\gtfn^g\simeq -i\left[R_\nu,U^{-1}t_a U\right]
\psi_+\psi_-$, we obtain the Noether current of the
left-transformation in the Skyrme model:
\begin{equation}
j_{L,a}^\mu(x)\simeq
2\kappa\tr\left\{\left(\frac{1}{\pi}L^\mu
+c_S\bigl[[L^\mu,L^\nu],L_\nu\bigr]\right)t_a\right\}
-\frac{i N_c}{48\pi^2}\epsilon^{\mu\nu\rho\lambda}\tr\bigl(
L_\nu L_\rho L_\lambda t_a\bigr) ,
\label{eq:j^La_mu_SK}
\end{equation}
with
\begin{equation}
L_\mu(x)=U(x)R_\mu(x)U(x)^{-1}=iU(x)\p_\mu U(x)^{-1} .
\end{equation}
Another way to get the same $j_{L,a}^\mu(x)$ in the low energy limit
is to repeat the arguments leading to \eqref{eq:j^Ra_mu_SK} by
replacing $g(x,z)^{-1}$ of \eqref{eq:Appg^-1} with
$g(x,z)^{-1}U(x)^{-1}\simeq \psi_-(z)+U(x)^{-1}\psi_+(z)$
which also realizes $\cA^{Ug}_z=0$.

Finally in this section, we comment on the low energy limit of another
candidate chiral current $\wt{j}_{L/R}^\mu(x)$ \eqref{eq:wtj} defined
on the boundary $z=\pm\infty$. This current should also be considered
under the boundary condition \eqref{eq:cA_M->0}.
Using the relation
$\cF_{\mu z}(x,z)=g(x,z)^{-1}\cF^g_{\mu z}(x,z)g(x,z)$
together with $g(x,z=-\infty)=\bm{1}$ and $g(x,z=\infty)=U(x)^{-1}$,
we find that the low energy limit of $\wt{j}_{L/R}^\mu(x)$ can
reproduce only the first term
$\pm(2\kappa/\pi)\tr\bigl((L/R)^\mu t_a\bigr)$ of the whole Noether
currents \eqref{eq:j^Ra_mu_SK} and \eqref{eq:j^La_mu_SK} of the Skyrme
model.

\section{Static properties of nucleons}
\label{sec:staticprop}

Baryons in the SS-model \eqref{eq:S=SYM+SCS} are described by a
soliton solution. In \cite{HSSY}, explicit construction of the baryon
solution and its collective coordinate quantization were given in the
two flavor ($N_f=2$) case and in the approximation of large 't\,Hooft
coupling $\lambda$. The baryon solution in this approximation on a
time slice is the BPST instanton solution \cite{BPST} with a fixed
size, which is determined by the balance between the contraction force
from the warp factors and the expansive one from the self-interaction
via the CS term. (See \cite{HMSU3} for an extension to the three
flavor case.)

In this section, we calculate the various static properties of
nucleons in the $N_f=2$ SS-model by using the chiral current
\eqref{eq:ourj=intourJ} and the baryon solution with quantized
collective coordinates given in \cite{HSSY}.
The quantities we calculate are
\begin{itemize}
\item
Electric charge distribution and charge radii

\item
Magnetic moments and magnetic charge radii

\item
Axial-vector coupling constant $g_A$.

\end{itemize}
The present analysis is an SS-model extension of that given in
\cite{ANW} for the Skyrme model. At each step, we compare our results
with the corresponding ones in the Skyrme model \cite{ANW} and
the experimental values.

\subsection{Baryon solution and its collective coordinate quantization}
\label{subsec:quantization}

In this subsection, we summarize the baryon solution in the $N_f=2$
SS-model and its collective coordinate quantization in the
approximation of large $\lambda$ \cite{HSSY}.
They are consistently given by assuming that the space-time coordinates
$x^M=(x^\mu,z)$ and the $U(2)$ gauge field $\cA_M$ are of the
following orders with respect to $\lambda (\gg 1)$:
\begin{align}
x^{M=i,z}&=\cO\left(\lambda^{-1/2}\right),\quad
x^0=\cO\left(\lambda^0\right)
\nn\\
\cA_{M=i,z}&=\cO\left(\lambda^{1/2}\right),\quad
\cA_0=\cO\left(\lambda^0\right)
\nn\\
\cF_{MN}&=\cO\left(\lambda\right),\quad
\cF_{0M}=\cO\left(\lambda^{1/2}\right),\quad
(M,N\ne 0) .
\label{eq:orders}
\end{align}
At the leading order in this large $\lambda$ approximation, the warp
factors \eqref{eq:h(z)k(z)} are set equal to $1$, and the $S_\YM$
\eqref{eq:SYM} is reduced to the YM action on the flat space.
The non-trivial effects of the warp factors as well as the CS term
\eqref{eq:SCS} are of order $\lambda^{-1}$.

In \cite{HSSY}, they expressed the action and the EOM in terms of the
rescaled coordinates and fields which are of order $\lambda^0$.
Here, we continue using the original coordinates and fields since
we need chiral currents as functions of the real space
coordinate.\footnote{
Accordingly, various constants in this paper (for example,
\eqref{eq:rho_st}, \eqref{eq:m} and \eqref{eq:Q}) differ from the
corresponding ones in \cite{HSSY} and \cite{HMSU3} by $\lambda$ or
$1/\lambda$.
}
Decomposing the $U(2)$ gauge field $\cA_M$ into the $SU(2)$ part $A_M$
and the $U(1)$ part $\Ah_M$ as
\begin{equation}
\cA_M=A_M + \Ah_M\,\frac12\bm{1}_2
=A^a_M\,t_a +\Ah_M\,\frac12\bm{1}_2,\quad
\left(t_a=\frac12\tau_a\right),
\end{equation}
the static baryon solution $\cA^\cl_M$ sitting at the origin
$(\bm{x},z)=(0,0)$ is given to the leading order by
\begin{align}
&A^\cl_{M=i,z}(\bm{x},z)
=-i\ol{f}(\xi)\,\ginst(\bm{x},z)^{-1}\p_M
\ginst(\bm{x},z),\quad
A^\cl_0(\bm{x},z)=0 ,
\label{eq:A^cl_SG}
\\
&\Ah^\cl_{M=i,z}(\bm{x},z)=0,\qquad
\Ah^\cl_0(\bm{x},z)=-\frac{1}{8\pi^2 a\lambda}\frac{1}{\xi^2}
\!\left[1-\frac{\rho^4}{(\xi^2 +\rho^2)^2}\right] ,
\label{eq:Ah^cl}
\end{align}
with
\begin{align}
\ginst(x)&=\frac{1}{\xi}\left(z \bm{1}_2+ix^i\tau_i\right)\in SU(2) ,
\\
\ol{f}(\xi)&= \frac{\rho^2}{\xi^2+\rho^2},\qquad
\xi=\sqrt{\bm{x}^2+z^2}\, .
\end{align}
The explicit expressions of the $SU(2)$ part of the gauge field and
the field strengths are
\begin{equation}
A^\cl_i(\bm{x},z)=\frac{2}{\xi^2}\ol{f}(\xi)\left(
z t_i-\epsilon_{ija}x^j t_a\right),\quad
A^\cl_z(\bm{x},z)=-\frac{2}{\xi^2}\ol{f}(\xi)x^a t_a ,
\label{eq:A^cl_M_SG_explicit}
\end{equation}
and
\begin{equation}
F^\cl_{ij}(\bm{x},z)=\frac{4}{\rho^2}\ol{f}(\xi)^2
\epsilon_{ija}\,\ginst^{-1}t_a\ginst,
\quad
F^\cl_{iz}(\bm{x},z)=\frac{4}{\rho^2}\ol{f}(\xi)^2
\ginst^{-1}t_i\ginst ,
\label{eq:F^cl_ij_iz}
\end{equation}
with
\begin{equation}
\ginst(\bm{x},z)^{-1}t_a\,\ginst(\bm{x},z)
=\frac{1}{\xi^2}\left[(z^2-\bm{x}^2)t_a
+2x^ax^b t_b-2\epsilon_{aib}z x^i t_b\right] .
\label{eq:ginst^-1t_aginst}
\end{equation}
The $SU(2)$ part of the solution is nothing but the BPST instanton
solution \cite{BPST} with size $\rho$ in the $(\bm{x},z)$ space .
The size $\rho$ is determined by minimizing the subleading part of the
energy as
\begin{equation}
\rhost^2=\frac{1}{8\pi^2 a\lambda}\sqrt{\frac65} .
\label{eq:rho_st}
\end{equation}
This implies that the baryon has a very small size of order
$\lambda^{-1/2}$. The mass of the solution with $\rho$ of
\eqref{eq:rho_st} is given by
\begin{equation}
M=8\pi^2\kappa +\sqrt{\frac{2}{15}}\,N_c ,
\label{eq:M}
\end{equation}
where the second subleading term is from the CS term and the $z^2$
terms of the warp factors.

The baryon solution we presented above is in the singular gauge,
namely, the gauge where $A^\cl_M$ is singular at the origin
$\xi=0$ but is regular at the infinity $\xi=\infty$.
Besides the singular gauge solution, we have the solution in the
regular gauge where $A^\cl_M$ is regular at the origin but is singular
at the infinity. The two are connected by the gauge transformation in
terms of $\ginst$. The reason why we adopt the singular gauge solution
here is related to the boundary condition \eqref{eq:cA_M->0} as we
will discuss in appendix \ref{app:regulargauge}.
In fact, the large $\xi$ behavior of the $SU(2)$ gauge field $A^\cl_M$
\eqref{eq:A^cl_SG} in the singular gauge is
$A^\cl_M\sim -i(\rho^2/\xi^2)\ginst^{-1}\p_M\ginst=\cO(1/\xi^3)$,
while it is $-i\ginst\p_M\ginst^{-1}=\cO(1/\xi)$ in the regular
gauge.

The collective coordinate quantization of this baryon solution is
carried out in a standard manner \cite{HSSY}. Besides the
center-of-mass $x^i$ coordinate and the $SU(2)$ rotation which are
genuine zero-modes, we take the size $\rho$ and the center-of-mass $z$
coordinate as approximate collective coordinates for quantization
since the energies of these modes are much lighter than other kinds of
massive modes for large $\lambda$.
Then, the gauge field one-form $\cA$ with the collective coordinates
of the center-of-mass and the size,
$X^\alpha(t)=\bigl(X^M(t),\rho(t)\bigr)$, and that of the $SU(2)$
rotation, $W(t)\in SU(2)$, incorporated is given by\footnote{
In this paper we adopt the way of introducing the collective
coordinate given in \cite{HMSU3}, which is related to that in
\cite{HSSY} by a gauge transformation.
}
\begin{equation}
\cA(\bm{x},z,t)=W(t)\left(\cA^\cl\bigl(\bm{x},z;X^\alpha(t)\bigr)
+\olPhi(\bm{x},z,t)\,dt-id\right)W(t)^{-1} ,
\label{eq:cA(bmx,z,t)}
\end{equation}
where $\olPhi$ is determined by the Gauss law as
\begin{equation}
\olPhi(\bm{x},z,t)=
\sum_{a=1}^3\chi^a(t)\olPhi_a-\dot{X}^M(t)
A^\cl_M(\bm{x},z;X^\alpha(t)) ,
\label{eq:olPhi}
\end{equation}
with
\begin{align}
\chi^a(t)&=-2i\tr\left(t_a W(t)^{-1}\dot{W}(t)\right) ,
\\
\olPhi_a&=f\bigl(\xi;X^\alpha(t)\bigr)t_a ,
\label{eq:olPhi_a}
\\
f(\xi)&=1-\ol{f}(\xi)=\frac{\xi^2}{\xi^2 +\rho^2} .
\label{eq:f(xi)}
\end{align}
In $\cA^\cl\bigl(\bm{x},z;X^\alpha(t)\bigr)$,
we must replace $x^i$, $z$ and $\rho$ in the initial expression
\eqref{eq:A^cl_SG} by $x^i-X^i(t)$, $z-Z(t)$ and $\rho(t)$,
respectively. This is the case also for $f\bigl(\xi;X^\alpha(t)\bigr)$
in \eqref{eq:olPhi_a}.
In later subsections, we use the following expressions of the gauge
field and the field strength which are derived from
\eqref{eq:cA(bmx,z,t)}:
\begin{align}
A_i(\bm{x},z,t)&=W(t)A^\cl_i\bigl(\bm{x},z;X^\alpha(t)\bigr)W(t)^{-1} ,
\label{eq:A_i=WA^cl_iW^-1}
\\
F_{MN}(\bm{x},z,t)&=W(t)F^\cl_{MN}\bigl(
\bm{x},z;X^\alpha(t)\bigr)W(t)^{-1} ,\quad
(M,N\ne 0) ,
\label{eq:F_MN=WF^cl_MNW^-1}
\\
F_{0M}(\bm{x},z,t)&=W(t)\left(
\dot{X}^i F^\cl_{Mi}+\dot{Z} F^\cl_{Mz}
+\dot{\rho}\Drv{}{\rho}A^\cl_M-\chi^a D^\cl_M\olPhi_a
\right)W(t)^{-1} .
\label{eq:F_0M}
\end{align}
Substituting the expressions of the fields in terms of the collective
coordinates into the original action \eqref{eq:S=SYM+SCS}, we obtain
the lagrangian of the collective coordinates:
\begin{equation}
L=L_X + L_Z + L_\rho + L_{\rho W} ,
\label{eq:Lcc}
\end{equation}
where the component lagrangians are given by
\begin{align}
L_X&=-8\pi^2\kappa +\frac{m_X}{2}\dot{\bm{X}}^2 ,
\\
L_Z&=\frac{m_Z}{2}\left(\dot{Z}^2-\omega_Z^2 Z^2\right) ,
\label{eq:L_Z}
\\
L_\rho&=\frac{m_\rho}{2}\left(\dot{\rho}^2-\omega_\rho^2\rho^2\right)
-\frac{Q}{\rho^2} ,
\\
L_{W\rho}&=\frac18 m_\rho^2\rho^2\sum_{a=1}^3(\chi^a)^2
=\cI(\rho)\tr\!\left[\bigl(-iW^{-1}\dot{W}\bigr)^2\right] ,
\label{eq:L_Wrho}
\end{align}
with the various quantities defined by
\begin{align}
m_X&=m_Z=\frac{m_\rho}{2}
=8\pi^2\kappa=\frac{\lambda N_c}{27\pi} ,
\label{eq:m}
\\
\omega_Z^2&=\frac23,\quad \omega_\rho^2=\frac16 ,
\\
Q&=\frac{27\pi N_c}{5\lambda} ,
\label{eq:Q}
\\
\cI(\rho)&=\frac14 m_\rho^2\rho^2=4\pi^2\kappa\rho^2 .
\end{align}

The collective coordinates $X^i(t)$, $Z(t)$, $\rho(t)$ and $W(t)$ are
quantized by using the lagrangian \eqref{eq:Lcc}.
In this paper, we consider only nucleons at rest and hence omit the
collective coordinate $X^i(t)$.
The quantization of other collective coordinates are summarized as
follows \cite{HSSY}:
\begin{itemize}
\item
Upon quantization of the $SU(2)$ rotation $W(t)$, baryons are
classified by the spin $J_i$ and the isospin $I_a$, which are the
Noether charges of the right and the left transformations on $W$,
$W\to g_I W g_J^{-1}$, respectively.
Explicitly, we have
\begin{align}
J_i&=2\cI(\rho)\tr\bigl(-iW^{-1}\dot{W}t_i\bigr)=\cI(\rho)\chi^i(t) ,
\label{eq:J_i}
\\
I_a&=2\cI(\rho)\tr\bigl(i\dot{W}W^{-1} t_a\bigr) .
\label{eq:I_a}
\end{align}
Since $J_i$ and $I_a$ are related by $I_at_a=-W J_it_i W^{-1}$, their
representations must be the same.

\item
For the $I=J=\ell/2$ state ($\ell$ is an odd integer), the wave
function $R_{\ell,n_\rho}(\rho)$ of $\rho$ (under the measure
$\int_0^\infty\!d\rho\,\rho^3$) is given by
\begin{equation}
R_{\ell,n_\rho}(\rho)=
\rho^{\wt{\ell}}F(-n_\rho,\wt{\ell}+2,m_\rho\omega_\rho\rho^2)
\exp\left(-\frac12 m_\rho\omega_\rho \rho^2\right),
\quad
(n_\rho=0,1,2,\ldots),
\label{eq:R(rho)}
\end{equation}
where $F(\alpha,\gamma;z)$ is the confluent hypergeometric function,
and $\wt{\ell}$ is related to $\ell$ by
$\wt{\ell}=-1+\sqrt{(\ell+1)^2+2 m_\rho Q}$.

\item
The lagrangian $L_Z$ \eqref{eq:L_Z} for $Z$ is simply that of a
harmonic oscillator. We denote its quantum number by $n_Z
(=0,1,2,\ldots)$.
\end{itemize}
Therefore, the baryon states are specified by a set of quantum numbers
$(\ell,n_\rho,n_Z)$. The mass of the corresponding state is given
in the $\MKK=1$ unit by
\begin{equation}
M_{\ell,n_\rho,n_Z}=8\pi^2\kappa
+\sqrt{\frac{(\ell+1)^2}{6}+\frac{2}{15}N_c^2}
\,+\sqrt{\frac23}\left(n_\rho+n_Z + 1\right) .
\end{equation}
The nucleon $N$ and $\Delta(1232)$ correspond to
$(\ell,n_\rho,n_Z)=(1,0,0)$ and $(3,0,0)$, respectively.
The mass of the nucleon and the $N$-$\Delta$ mass difference are
\begin{align}
M_N&=8\pi^2\kappa+\sqrt{\frac23+\frac{2}{15}N_c^2}
+\sqrt{\frac23} ,
\label{eq:M_Norg}
\\
M_\Delta-M_N&=\sqrt{\frac83+\frac{2}{15}N_c^2}\,
-\sqrt{\frac23+\frac{2}{15}N_c^2} .
\label{eq:M_D-M_N}
\end{align}

In the calculations of physical quantities in the following
subsections, we make the following treatments on $\MKK$, $M_N$, $\rho$
and $Z$:
\begin{itemize}
\item
We determine the value of $\MKK$ by equating the $N$-$\Delta$ mass
difference \eqref{eq:M_D-M_N} with $N_c=3$ with its experimental
value, $(1232-939)\,\MeV$:
\begin{equation}
\MKK=488\,\MeV, \quad
1/\MKK=0.404\,\fm .
\label{eq:MKK=}
\end{equation}

\item
Since our analyses below on currents are at the leading order in
$1/\lambda$, we take only the first term $8\pi^2\kappa$ of the
nucleon mass \eqref{eq:M_Norg}:
\begin{equation}
M_N=8\pi^2\kappa .
\label{eq:M_N=8pi^2kappa}
\end{equation}

\item
Since we have $F(0,\wt{\ell}+2,m_\rho\omega_\rho\rho^2)\equiv 1$,
the wave function \eqref{eq:R(rho)} of $\rho$ is given for nucleons by
\begin{equation}
R_{1,0}(\rho)=\rho^\ellN
\exp\left(-\frac12 m_\rho\omega_\rho \rho^2\right),
\quad
\left(\ellN=-1+2\sqrt{1+N_c^2/5}\right).
\label{eq:R(rho)_N}
\end{equation}
In evaluating the various quantities $\cO(\rho)$ depending on $\rho$,
we take the expectation value using the wave function
\eqref{eq:R(rho)_N} and the measure $\int_0^\infty\!d\rho\rho^3$:
\begin{equation}
\rhoVEV{\cO(\rho)}=\frac{
\int_0^\infty\!d\rho\,\rho^3 \cO(\rho)R_{1,0}(\rho)^2}{
\int_0^\infty\!d\rho\,\rho^3 R_{1,0}(\rho)^2} .
\end{equation}
In particular, $\VEV{\rho^2}_\rho$, which repeatedly appears in the
following subsections, is given by
\begin{equation}
\rhoVEV{\rho^2}=\frac{2+\ellN}{m_\rho\omega_\rho}
=\frac{\sqrt{6}}{8\pi^2\kappa}\left(
\frac12+\sqrt{1+\frac{N_c^2}{5}}\right) .
\label{eq:VEVrho^2}
\end{equation}
This agrees with $\rhost^2$ \eqref{eq:rho_st} in the large $N_c$
limit. Using \eqref{eq:M_N=8pi^2kappa} and
\begin{equation}
M_N=939\,\MeV=1.92\,\MKK ,
\label{eq:M_NbyM_KK}
\end{equation}
the numerical value of \eqref{eq:VEVrho^2} for $N_c=3$ is
\begin{equation}
\sqrt{\rhoVEV{\rho^2}}=0.672\,\fm ,
\label{eq:sqrtVEVrho^2_num}
\end{equation}
where we have used \eqref{eq:MKK=} for expressing the result in
$\fm$ unit.

\item
Since the $Z(t)$ dependence in the five-dimensional current of the
form $z-Z(t)$ disappears after the $z$-integration necessary in
obtaining the four-dimensional current, we put $Z=0$ from the start.
The time-derivative terms of $Z$ appearing in our chiral
current is only the $\dot{Z}$ term coming from \eqref{eq:F_0M}, and no
terms containing more time-derivatives arise. Since the expectation
value of $\dot{Z}$ is equal to zero for energy eigenstates of the $Z$
harmonic oscillator, we drop this $\dot{Z}$ term in the current.

\end{itemize}

In the following subsections we calculate the various static
properties of nucleons using the $U(N_f)_{L/R}$ currents defined
by \eqref{eq:ourj=intourJ} and \eqref{eq:ourJ_mu}. We see from
\eqref{eq:F^cl_ij_iz} and \eqref{eq:F_0M} that another chiral
current $\wt{j}^\mu_{L/R}(x)$ \eqref{eq:wtj} defined by the field
strength on the boundary $z=\pm\infty$ does vanish. This is the case
both in the singular gauge adopted here and also in the regular
gauge.
Quite similarly, the $z$-component of the five-dimensional current,
\eqref{eq:ourJ_z}, vanishes on the boundary for the baryon
configuration and hence satisfies the condition \eqref{eq:cJ_z->0}.

Although the local current $\wt{j}^\mu_{L/R}(x)$ itself vanishes
for the baryon configuration, it may happen that the space integration
of a quantity containing the current is non-vanishing if we take the
$z\to\pm\infty$ limit after carrying out the integration. This is the
case for the isovector magnetic moment computed in sec.\
\ref{subsec:magneticmoments}.
We will briefly summarize in sec.\ \ref{subsec:compwtj} the results of
computation using $\wt{j}^\mu_{L/R}(x)$.

\subsection{Charge density}
\label{subsec:chargedensity}

As a static property of nucleons, let us first consider its charge
density.
For this we need the isospin density $j_{V,a}^0(\bm{x})$ and
the baryon number density. First, the isospin density before
the $z$-integration, $\fJ_{V,a}^0(x,z)$ ($a=1,2,3$), to the leading
order in the large $\lambda$ approximation of \eqref{eq:orders} is
obtained from the non-abelian part of \eqref{eq:ourJ_mu} by neglecting
the warp factors and using that $\psi_V(z)=1$ for the vector current
(see \eqref{eq:psi_VA}):
\begin{align}
\fJ_{V,a}^0(x,z)&=2i\kappa\tr\Bigl\{\Bigl(
\left[F_{0i},A_i\right]
+\left[F_{0z},A_z\right]\Bigr)t_a\Bigr\}
\nn\\
&=2\kappa\tr\Biggl\{\Biggl(
\sum_{M=i,z}i\left[\Drv{}{\rho}A^\cl_M,A^\cl_M\right]\dot{\rho}(t)
-\sum_{M=i,z}i\bigl[D^\cl_M\olPhi_b,A^\cl_M\bigr]\chi^b(t)
\Biggr)W(t)^{-1}t_a W(t)\Biggr\} ,
\label{eq:J_V,a^0}
\end{align}
where have dropped the $\dot{X}^i$ and $\dot{Z}$ terms in
\eqref{eq:F_0M} as announced above. Using
\begin{equation}
\sum_{M=i,z}i\bigl[D^\cl_M\olPhi_a,A^\cl_M\bigr]
=\frac{8}{\rho^2}\ol{f}(\xi)^3\,t_a ,
\label{eq:sum_M[DolPhi,A]=}
\end{equation}
and that $\left[(\p/\p\rho)A^\cl_M,A^\cl_M\right]=0$, which is due to
$(\p/\p\rho)A^\cl_M\propto A^\cl_M$, we get
\begin{equation}
\fJ_{V,a}^0(x,z)=\frac{2}{\pi^2}
\frac{\rho^2}{(\xi^2+\rho^2)^3}\,I_a ,
\label{eq:J_V,a^0==}
\end{equation}
where $I_a$ is the isospin operator \eqref{eq:I_a}, and we have used
the formula
$\tr\bigl(t_bW^{-1}t_a W\bigr)\chi^b
=\tr\bigl(-i\dot{W}W^{-1}t_a\bigr)$.
The isospin density in four dimensions is obtained by integrating
\eqref{eq:J_V,a^0==} over $z$:
\begin{equation}
j_{V,a}^0(t,\bm{x})=\int_{-\infty}^\infty\!dz\,\fJ_{V,a}^0(x,z)
=\frac{3}{4\pi}\frac{\rho^2}{(r^2+\rho^2)^{5/2}}\,I_a .
\label{eq:j^Va_0}
\end{equation}
We can confirm that our $j_{V,a}^0$ is indeed the isospin density:
\begin{equation}
I_a=\int\! d^3x\,j_{V,a}^0(t,\bm{x}) .
\end{equation}

Next is the baryon number density in four dimensions.
In the SS-model, a (topologically) conserved baryon number current in
five dimensions has been defined \cite{SaSu1,SaSu2}:
\begin{equation}
J_B^M(x,z)=\frac{1}{32\pi^2}\epsilon^{MNPQR}
\tr\bigl(\cF_{NP}\cF_{QR}\bigr) .
\label{eq:J_B^M}
\end{equation}
Our baryon solution has in fact a unit baryon number; namely, we have
$\int\! d^3x dz J_B^0=1$ for the solution \eqref{eq:A^cl_SG}.
It is interesting to notice that this topological current $J_B^M$
\eqref{eq:J_B^M} is obtained as the Noether current of the $U(1)_V$
symmetry, $\fJ_\gtfn^M$ \eqref{eq:cJ^zeta_M} with $\gtfn=\bm{1}$
(see also \eqref{eq:ourJ_mu}), divided by $N_c$, though all the
fields in the action \eqref{eq:S=SYM+SCS} are inert under the $U(1)_V$
transformation.
{}From \eqref{eq:J_B^M}, the four-dimensional baryon number density
for the baryon configuration \eqref{eq:F_MN=WF^cl_MNW^-1} is
\begin{equation}
j_B^0(t,\bm{x})=\int_{-\infty}^\infty\! dz J_B^0(x,z)
=\frac{1}{8\pi^2}\int_{-\infty}^\infty\! dz\,\epsilon_{ijk}
\tr\bigl(F^\cl_{ij} F^\cl_{kz}\bigr)
=\frac{15}{8\pi}\frac{\rho^4}{(r^2+\rho^2)^{7/2}} .
\label{eq:j_B^0}
\end{equation}
Integration of \eqref{eq:j_B^0} further with respect to $\bm{x}$ gives
one as mentioned above.

\begin{figure}[hptb]
\centering
\epsfxsize=8.0cm
\epsfbox{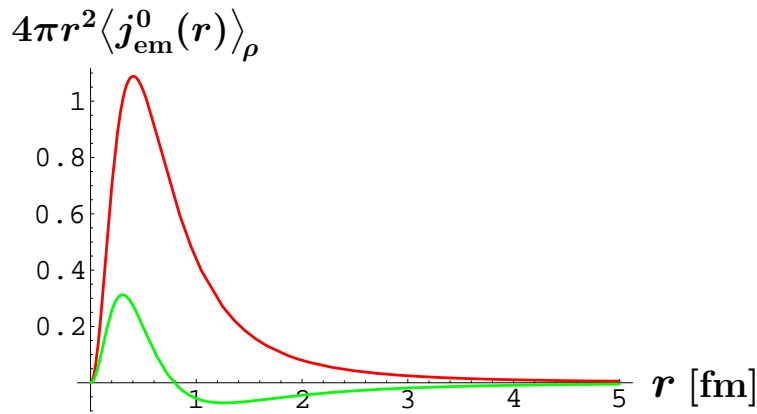}
\put(7,55){{\Large $\bm{r}$} {\large$\bm{[\fm]}$}}
\put(-235,190){{\large $\bm{4\pi r^2\!\rhoVEV{j_\elmg^0(r)}}$}}
\vspace{-12mm}
\caption{
Radial charge distribution $4\pi r^2 \!\rhoVEV{j_\elmg^0(r)}$ of proton
(red curve) and neutron (green curve). The units of the horizontal and
the vertical axes are $\fm$ and $1/\fm$, respectively.
}
\label{fig:rhoem}
\end{figure}

Now, let us evaluate physical quantities from the isospin density
\eqref{eq:j^Va_0} and the baryon number density \eqref{eq:j_B^0}.
First, the charge density is given by
\begin{equation}
j_\elmg^0(t,\bm{x})=j_{V,a=3}^0(t,\bm{x})+\frac12 j_B^0(t,\bm{x})
=\frac{3}{4\pi}\frac{\rho^2}{(r^2+\rho^2)^{5/2}}\,I_3
+\frac{15}{16\pi}\frac{\rho^4}{(r^2+\rho^2)^{7/2}} ,
\label{eq:j^em_0}
\end{equation}
and we actually have $\int\! d^3x\,j_\elmg^0(\bm{x})=I_3 +(1/2)$.
In fig.\ \ref{fig:rhoem}, we plot the radial charge distribution
$4\pi r^2\!\rhoVEV{j_\elmg^0(r)}$ for proton and neutron.
The curves in fig.\ \ref{fig:rhoem} are close to those in the Skyrme
model (see fig.\ 2 in \cite{ANW}). However, they are in disagreement
with experimental results \cite{NQ}, which show that the neutron is
almost locally neutral in the region $r\gtrsim0.05\,\fm$, and that the
charge distribution of proton decays exponentially for large $r$
(while our $4\pi r^2 \!\rhoVEV{j_\elmg^0(r)}$ decays as $1/r^5$).

Next, the isoscalar mean square charge radius is calculated
as\footnote{
$\VEV{r^2}_{I=0}^{1/2}$ as a function of $\rho^2$ was obtained before
by S.\ Sugimoto (private communication).
}
\begin{equation}
\sqrt{\VEV{r^2}_{I=0}}=\sqrt{\rhoVEV{
\int\! d^3x\,r^2 j_B^0(t,\bm{x})}}
=\sqrt{\frac32\rhoVEV{\rho^2}}=0.82\,\fm ,
\quad
\left(\ANW: 0.59\,\fm,\ \Exp: 0.80\,\fm\right),
\label{eq:<r^2>_I=0}
\end{equation}
where we have used \eqref{eq:sqrtVEVrho^2_num}, and $\ANW$ and $\Exp$
denote the result of \cite{ANW} and the experimental value \cite{PDG},
respectively. The (numerator of) isovector mean square charge radius,
\begin{equation}
\VEV{r^2}_{I=1}=\frac{
\rhoVEV{\int\! d^3x\,r^2 j_{V,a=3}^0(t,\bm{x})}}{
\rhoVEV{\int\! d^3x\,j_{V,a=3}^0(t,\bm{x})}} ,
\end{equation}
is logarithmically divergent at $r=\infty$ as in the Skyrme model
\cite{ANW}. This phenomenon may be ascribed to the fact that all the
quarks and hence the pions are massless in the SS-model
\cite{BegZepeda}.

\subsection{Magnetic moments}
\label{subsec:magneticmoments}

In this subsection, we examine quantities related with the
magnetic moment of nucleons:
\begin{equation}
\bm{\mu}=\frac12\int\!d^3 x\,\bm{x}\times\bm{j}_\elmg(\bm{x})
=\frac12\bm{\mu}_{I=1}+\frac12\bm{\mu}_{I=0} ,
\end{equation}
where $\bm{j}^\elmg$ is the electro-magnetic current vector,
\begin{equation}
\bm{j}_\elmg=\bm{j}_{V,a=3} +\frac12\,\bm{j}_B ,
\end{equation}
and the isovector and the isoscalar magnetic moments are defined
respectively by\footnote{
Our $SU(N_f)_{V/A}$ current $j_{V/A,a}^i$ is half of that in
\cite{ANW}.
}
\begin{align}
\bm{\mu}_{I=1}&=\int\!d^3 x\,\bm{x}\times\bm{j}_{V,a=3}(\bm{x}),
\label{eq:bmmu_I=1}
\\
\bm{\mu}_{I=0}&=\frac12\int\!d^3 x\,\bm{x}\times\bm{j}_B(\bm{x}) .
\label{eq:bmmu_I=0}
\end{align}
For this purpose we need the space components of the isospin current
and the baryon number current. The space component of the five
dimensional $SU(N_f)_{V/A}$ currents at the leading order in the
large $\lambda$ approximation are given from \eqref{eq:ourJ_mu}
and \eqref{eq:orders} by
\begin{equation}
\fJ_{V/A,a}^i(x,z)=-2\kappa\tr\left\{\left(
\sum_{M=j,z}i\left[F^\cl_{iM},A^\cl_M\right]\psi_{V/A}(z)
+F^\cl_{iz}\drv{\psi_{V/A}(z)}{z}\right)W(t)^{-1}t_aW(t)\right\}.
\label{eq:J_V/A,a^i}
\end{equation}
Here, we need the vector current. Using $\psi_V(z)=1$ and
\begin{equation}
\sum_{M=j,z}i\left[F^\cl_{iM},A^\cl_M\right]
=\frac{16\rho^4}{\xi^2(\xi^2+\rho^2)^3}\left(
z t_i-\epsilon_{ija}x^j t_a\right) ,
\label{eq:[F^cl_iM,A^cl_M]}
\end{equation}
the space component of the four-dimensional $SU(N_f)_V$ current is
obtained as
\begin{equation}
j_{V,a}^i(t,\bm{x})=\!\int_{-\infty}^\infty\! dz\,\fJ_{V,a}^i(x,z)
=\frac{4\pi\kappa}{\rho^2}\left(\frac{8}{r}
-\frac{8 r^4+ 20\rho^2 r^2 + 15 \rho^4}{(r^2+\rho^2)^{5/2}}
\right)\epsilon_{ijk} x^j\tr\bigl(t_k W(t)^{-1}t_a W(t)\bigr) .
\label{eq:j^Va_i}
\end{equation}
As for the space component of the baryon number current, we first have
from \eqref{eq:J_B^M}, \eqref{eq:F_MN=WF^cl_MNW^-1} and
\eqref{eq:F_0M} that
\begin{align}
\fJ_B^i(x,z)&=-\frac{1}{8\pi^2}\,\epsilon_{ijk}
\tr\!\left(F_{jk}F_{0z}+2 F_{0j} F_{kz}\right)
\nn\\
&=\frac{3}{\pi^2}\frac{\rho^4}{(\xi^2 +\rho^2)^4}
\left[\left(\delta_{ia}-\epsilon_{ija}x^j\right)\chi^a(t)
+ 2x^i\drv{}{t}\ln\rho(t)\right] ,
\end{align}
and hence
\begin{equation}
j_B^i(t,\bm{x})=\int_{-\infty}^\infty\! dz\,\fJ_B^i(x,z)
=\frac{15}{16\pi}\frac{\rho^4}{(r^2+\rho^2)^{7/2}}
\left(-\epsilon_{ija}x^j \chi^a(t)
+2 x^i\drv{}{t}\ln \rho(t)\right) .
\label{eq:j^B_i}
\end{equation}

Then, the isovector and isoscalar magnetic moments,
\eqref{eq:bmmu_I=1} and \eqref{eq:bmmu_I=0}, are calculated as
follows:
\begin{align}
\left(\bm{\mu}_{I=1}\right)_i
&=\epsilon_{ijk}\int\! d^3 x\,x^j j_{V,a=3}^k(t,\bm{x})
=-8\pi^2\kappa\rho^2\tr\bigl(t_i W(t)^{-1}t_3 W(t)\bigr) ,
\label{eq:bmmu_I=1=}
\\
\left(\bm{\mu}_{I=0}\right)_i
&=\frac12\epsilon_{ijk}\int\! d^3 x\,x^j j_B^k(t,\bm{x})
=\frac{\rho^2}{4}\chi^i(t)=\frac{1}{16\pi^2\kappa}\,J_i ,
\label{eq:bmmu_I=0=}
\end{align}
where $J_i$ is the spin operator \eqref{eq:J_i}.
{}From this we can read off the isovector $g$-factor $g_{I=1}$ and the
isoscalar one $g_{I=0}$ defined for nucleon states in the Pauli
matrix representation of spin and isospin, $J_i=\sigma_i/2$ and
$I_a=\tau_a/2$, by
\begin{align}
\bm{\mu}_{I=1}&=\frac{g_{I=1}}{2 M_N}
\frac{\bm{\sigma}}{2}\otimes\tau_3 ,
\\
\bm{\mu}_{I=0}&=\frac{g_{I=0}}{2 M_N}\frac{\bm{\sigma}}{2} .
\end{align}
For $g_{I=1}$ we use the relation valid for nucleon states (see eq.\
(22) in \cite{ANW}),
\begin{equation}
\bra{N'}\tr\bigl(t_i W^{-1}t_a W\bigr)\ket{N}
=-\frac16\bra{N'}\sigma_i\otimes\tau_a\ket{N} ,
\label{eq:trtW^-1tW=}
\end{equation}
to get
\begin{equation}
g_{I=1}=\frac{16\pi^2\kappa}{3}M_N\rhoVEV{\rho^2}
=\sqrt{\frac23}\left(1+2\sqrt{1+\frac{N_c^2}{5}}\right)M_N
=6.83,
\quad
\left(\ANW: 6.38,\ \Exp: 9.41\right),
\label{eq:g_I=1=}
\end{equation}
where we have used \eqref{eq:VEVrho^2} and \eqref{eq:M_NbyM_KK}.
On the other hand, $g_{I=0}$ is exactly equal to one if we adopt
\eqref{eq:M_N=8pi^2kappa} for $M_N$:
\begin{equation}
g_{I=0}=\frac{M_N}{8\pi^2\kappa}=1 ,
\quad
\left(\ANW: 1.11,\ \Exp: 1.76\right) .
\label{eq:g_I=0=1}
\end{equation}
The results \eqref{eq:g_I=1=} and \eqref{eq:g_I=0=1} are restated
into the nucleon magnetic moments $\mu_{p,n}$ in units of Bohr
magneton $e\hbar/(2M_N)$ as follows:
\begin{align}
\mu_p&=\frac14\left(g_{I=0}+g_{I=1}\right)=1.96 ,
\quad
\left(\ANW: 1.87,\ \Exp: 2.79\right) ,
\label{eq:mup}
\\
\mu_n&=\frac14\left(g_{I=0}-g_{I=1}\right)=-1.46 ,
\quad
\left(\ANW: -1.31,\ \Exp: -1.91\right) .
\label{eq:mun}
\end{align}
Their ratio is
\begin{equation}
\abs{\frac{\mu_p}{\mu_n}}=1.34 ,
\quad
\left(\ANW: 1.43,\ \Exp: 1.46\right) .
\label{eq:muo/mun}
\end{equation}

Finally, the isoscalar magnetic mean square radius is defined by
\begin{equation}
\VEV{r^2}_{M,I=0}=\frac{
\rhoVEV{\int_0^\infty\!dr\,r^2\rho_{M,I=0}(r)}}{
\rhoVEV{\int_0^\infty\!dr\,\rho_{M,I=0}(r)}} ,
\end{equation}
in terms of the isoscalar magnetic moment density $\rho_{M,I=0}(r)$
giving the coefficient of $\chi^i(t)$ in \eqref{eq:bmmu_I=0=};
$\left(\bm{\mu}_{I=0}\right)_i
=\int_0^\infty\!dr\,\rho_{M,I=0}(r)\,\chi^i(t)$.
In the present case, we have
\begin{equation}
\rho_{M,I=0}(r)=\frac54\frac{\rho^4r^4}{(r^2+\rho^2)^{7/2}} .
\end{equation}
Therefore, $\VEV{r^2}_{M,I=0}$ is logarithmically divergent at
$r=\infty$ in contrast with the case of the Skyrme model where it is
finite \cite{ANW}.

\subsection{Axial-vector coupling}
\label{subsec:g_A}

Let us consider the $SU(N_f)_A$ current to obtain the axial-vector
coupling $g_A$. First, (the space component of) the five-dimensional
current $\fJ_{A,a}^i(x,z)$ is given from \eqref{eq:J_V/A,a^i} by using
\eqref{eq:[F^cl_iM,A^cl_M]}, \eqref{eq:F^cl_ij_iz} and
\eqref{eq:ginst^-1t_aginst} for the baryon solution:
\begin{equation}
\fJ_{A,a}^i(x,z)=-\frac{8\kappa\rho^2}{\xi^2(\xi^2+\rho^2)^2}
\left\{
\frac{4\rho^2z\psi_A(z)}{\xi^2+\rho^2}
+\left(z^2-\frac{\bm{x}^2}{3}\right)\drv{\psi_A(z)}{z}\right\}
\tr\bigl(t_i W(t)^{-1}t_a W(t)\bigr) ,
\label{eq:J_A,a^i}
\end{equation}
where we have kept only the terms which are even in $z$ ($\psi_A(z)$
is odd in $z$) since we eventually carry out the $z$-integration, and
have carried out the spatial angle averaging to replace $x^a x^b$ in
\eqref{eq:ginst^-1t_aginst} by $\delta_{ab}r^2/3$.
Since we are considering the leading order of
the large $\lambda$ approximation with the orders of the respective
quantities given by \eqref{eq:orders} and we have already approximated
the warp factors by $1$ in \eqref{eq:J_A,a^i}, the function
$\psi_A(z)=(2/\pi)\arctan z$ \eqref{eq:psi_VA} should consistently
be approximated in \eqref{eq:J_A,a^i} by
\begin{equation}
\psi_A(z)=\frac{2}{\pi}\,z .
\label{eq:App_psi_A}
\end{equation}
We will comment later on what happens to the
$SU(N_f)_A$  current if we keep the original expression of the
function $\psi_A(z)$. Using \eqref{eq:App_psi_A}, the five-dimensional
$SU(N_f)_A$ current reads
\begin{equation}
j_{A,a}^i(t,\bm{x})=\int_{-\infty}^\infty\! dz\,\fJ_{A,a}^i(x,z)
=\frac{32\kappa}{3\rho^2}\left[8r-
\frac{8 r^4+ 12\rho^2 r^2 + 3\rho^4}{(r^2+\rho^2)^{3/2}}
\right]\tr\bigl(t_i W(t)^{-1}t_a W(t)\bigr) .
\label{eq:j_A,a^i}
\end{equation}
The axial-vector coupling constant $g_A$ is obtained by identifying
the nucleon matrix element of
\begin{equation}
\int\! d^3x\,j_{A,a}^i(\bm{x},t)
=-\frac{32\pi\kappa\rho^2}{3}\tr\bigl(t_i W^{-1}t_a W\bigr) ,
\label{eq:intj_A,a^i}
\end{equation}
with the $\bm{q}\to 0$ limit of the non-relativistic expression
\cite{ANW}:
\begin{equation}
\bra{N'(p')}j_{A,a}^i(0)\ket{N(p)}=\frac12 g_A(\bm{q}^2)
\left(\delta^{ij}-\frac{q^i q^j}{\bm{q}^2}\right)
\bra{N'}\sigma_j\otimes\tau_a\ket{N} ,
\quad \left(\bm{q}=\bm{p}'-\bm{p}\right) .
\label{eq:<N|j_A|N>}
\end{equation}
Making the angle averaging of $\bm{q}$ to replace $q^i q^j/\bm{q}^2$
in \eqref{eq:<N|j_A|N>} by $\delta^{ij}/3$ and using
\eqref{eq:trtW^-1tW=}, we get\footnote{
Eq.\ \eqref{eq:g_A=} before taking the expectation value of $\rho^2$
agrees with $g_{A,mag}=4C/\pi$
given in eq.\ (5.35) of \cite{HRYY2} using a different
approach. The constant $C$ is given below eq.\ (5.17) of
\cite{HRYY2}.
}
\begin{equation}
g_A=g_A(0)=\frac{16\pi\kappa}{3}\rhoVEV{\rho^2}
=\frac{\sqrt{6}}{3\pi}\left(1+2\sqrt{1+\frac{N_c^2}{5}}\right)
=1.13,
\quad
\left(\ANW: 0.61,\ \Exp: 1.24\right) ,
\label{eq:g_A=}
\end{equation}
where we have used \eqref{eq:VEVrho^2} for $\rhoVEV{\rho^2}$.
Compared with the result in the Skyrme model,
we have got a surprisingly good agreement with the experimental
value. Note that $g_A$ \eqref{eq:g_A=} is independent of our choice of
$\MKK$ \eqref{eq:MKK=}.
If we had simply replaced $\rhoVEV{\rho^2}$ in \eqref{eq:g_A=} by the
minimum of the potential $\rhost^2$ given by \eqref{eq:rho_st} without
treating $\rho$ as an approximate collective coordinate, we would have
obtained, instead of \eqref{eq:g_A=}, a less attractive result;
$g_A=2^{3/2}N_c/(\sqrt{15}\,\pi)=0.70$. This is nothing but the large
$N_c$ limit of \eqref{eq:g_A=}.

We will make some comments on the evaluation of $g_A$, which has been
done by the four-dimensional integration $\int\!d^3x\int\!dz$ of
the five-dimensional current \eqref{eq:J_A,a^i}.
First, note that the four-dimensional integration of the
$(z^2-\bm{x}^2/3)d\psi_A(z)/dz$ part is superficially logarithmically
divergent at $\xi=\infty$ for $\psi_A(z)$ of \eqref{eq:App_psi_A}.
However, it can actually be convergent since the sum of the
coefficients of $z^2$ and $(x^i)^2$ is equal to zero,
$1-(1/3)\times 3=0$, and it can take any value depending the way of
integration.
The integration vanishes if we adopt the symmetric integration using
the four-dimensional polar coordinates. Our $g_A$ \eqref{eq:g_A=} has
been obtained by carrying out first the $z$-integration to give
$j_{A,a}^i$ and then the $\bm{x}$-integration to take the
zero-momentum limit. If we reverse the order of integrations, we get
\begin{equation}
\int\!dz\int\!d^3x\,\fJ_{A,a}^i(x,z)=0 .
\label{eq:intdzintd^3xJ_A,a^i=0}
\end{equation}
This is seen from the following formula valid for any $\psi_A(z)$:
\begin{equation}
\int\! d^3x\,\fJ_{A,a}^i(x,z)=\frac{8\pi^2\kappa}{3\rho^2}
\drv{}{z}\!\Bigl[Q(z)\psi_A(z)\Bigr]
\tr\bigl(t_i W^{-1}t_a W\bigr) ,
\label{eq:intd^3xJ_A,a^i}
\end{equation}
with
\begin{equation}
Q(z)=8\abs{z}^3-\frac{8z^4+4\rho^2 z^2-\rho^4}{\sqrt{z^2+\rho^2}}
=\cO\!\left(\frac{1}{z^3}\right),\quad
\left(z\to\pm\infty\right) .
\end{equation}

Eq.\ \eqref{eq:intd^3xJ_A,a^i} implies that we have
\eqref{eq:intdzintd^3xJ_A,a^i=0} also for the original
expression $\psi_A(z)=(2/\pi)\arctan z$ \eqref{eq:psi_VA}.
Since this four-dimensional integration is absolutely and uniformly
convergent, we can freely exchange the order of integrations to
conclude that $g_A=0$ for the original $\psi_A(z)$ \eqref{eq:psi_VA}
and the trivial warp factors.

\subsection{Computations using $\bm{\wt{j}_{L/R}^\mu}$}
\label{subsec:compwtj}

In this subsection, we briefly summarize the computations using
the chiral current $\wt{j}_{L/R,a}^\mu(x)$ \eqref{eq:wtj} of the
bulk-boundary correspondence.
As mentioned at the end of sec.\ \ref{subsec:quantization}, although
the local current $\wt{j}_{L/R,a}^\mu(x)$ itself vanishes for the
baryon configuration, the space integration of a quantity containing
$\wt{j}_{L/R,a}^\mu(x)$ can be non-vanishing if we take the
$z\to\infty$ limit after carrying out the space integration.
Approximating the warp factor $k(z)$ by $1$ as before and replacing
the boundaries $z=\pm\infty$ by $z=\pm\Lambda$, the space
component of the $SU(N_f)_{L/R}$ current \eqref{eq:wtj} is given by
\begin{align}
\wt{j}_{L/R,a}^i(x)
&=\mp2\kappa\tr\bigl(F^{iz}(x,z=\pm\Lambda)\,t_a\bigr)
\nn\\
&=\mp\frac{8\kappa\rho^2}{\xi^2(\xi^2+\rho^2)^2}
\tr\Bigl\{\left[\bigl(z^2-\bm{x}^2\bigr)t_i+2x^ix^jt_j
-2\epsilon_{ijk}z x^j t_k\right]W^{-1}t_aW\Bigr\}
\Bigr|_{z=\pm\Lambda} ,
\end{align}
where we have used \eqref{eq:F^cl_ij_iz} and
\eqref{eq:ginst^-1t_aginst} for the singular gauge field strength.
Therefore, the space components of the vector and the axial-vector
currents are
\begin{align}
\wt{j}_{V,a}^i(x)&=\frac{32\kappa\rho^2\Lambda}{(r^2+\Lambda^2)^3}
\epsilon_{ijk}x^j\tr\bigl(t_kW^{-1}t_aW\bigr) ,
\label{eq:wtj_Va^i}
\\
\wt{j}_{A,a}^i(x)&=-\frac{16\kappa\rho^2}{(r^2+\Lambda^2)^3}
\left(\Lambda^2-\frac{r^2}{3}\right)\tr\bigl(t_iW^{-1}t_a W\bigr) ,
\label{eq:wtj_Aa^i}
\end{align}
where we have ignored $\rho^2$ in the denominators since we take the
limit $\Lambda\to\infty$ in the end.

First, the isovector magnetic moment is calculated using
\eqref{eq:wtj_Va^i} as
\begin{equation}
\left(\wt{\bm{\mu}}_{I=1}\right)_i
=\lim_{\Lambda\to\infty}
\epsilon_{ijk}\int\! d^3 x\,x^j\,\wt{j}_{V,a=3}^k(t,\bm{x})
=-16\pi^2\kappa\rho^2\tr\bigl(t_i W(t)^{-1}t_3 W(t)\bigr) .
\label{eq:wtbmmu_I=1=}
\end{equation}
This is twice our previous result \eqref{eq:bmmu_I=1=}. Therefore,
the corresponding isovector $g$-factor is
\begin{equation}
\wt{g}_{I=1}=2 g_{I=1}=13.6 ,
\label{eq:wtg_I=1}
\end{equation}
which should be compared with the values in \eqref{eq:g_I=1=}.
The factor of two difference between the two isovector magnetic
moments, \eqref{eq:bmmu_I=1=} and \eqref{eq:wtbmmu_I=1=}, can be
understood from the relation \eqref{eq:relation} between the currents.
In the present case, we have to cutoff the $z$-integrations for
$j_\gtfn^\mu(x)$ \eqref{eq:j^zeta_mu} and $\chi^{\mu\nu}(x)$
\eqref{eq:chi^munu} at $z=\pm\Lambda$.
One can confirm first that $\bm{\mu}_{I=1}$ \eqref{eq:bmmu_I=1=}
remains the same even if we take the $\Lambda\to\infty$ limit at the
end, namely, even if we exchange the order of the $\bm{x}$ and $z$
integrations. Second, we can show that the contribution of the
$\p_j\chi^{ij}$ term of \eqref{eq:relation} to the isovector magnetic
moment, $\epsilon_{ijk}\int\!d^3x\,\chi^{jk}$, is just what is
necessary to explain the factor of two difference.

Next, the axial-vector coupling $g_A$ becomes zero if we adopt
the axial-vector current of \eqref{eq:wtj_Aa^i}:
\begin{equation}
\lim_{\Lambda\to\infty}
\int\! d^3x\,\wt{j}_{A,a}^i(\bm{x},t)=0 .
\label{eq:intwtj_A,a^i=0}
\end{equation}
This is also consistent with the relation \eqref{eq:relation}.
In this case, the space integration of $\p_j\chi^{ij}$ vanishes, and
\eqref{eq:intwtj_A,a^i=0} is equivalent to our previous
\eqref{eq:intdzintd^3xJ_A,a^i=0}.

\section{Summary and discussions}
\label{sec:summary}

In this paper, we computed static properties of nucleons in the
SS-model, namely, the five-dimensional YM+CS theory realized as
the low energy effective theory of mesons in the holographic QCD model
of Sakai and Sugimoto \cite{SaSu1,SaSu2}.
For this purpose, we first constructed a chiral current in
four dimensions from the Noether current of local gauge transformation
which effects non-trivial transformations on the boundaries of the
extra fifth dimension. We confirmed that our chiral current
is reduced in the low energy limit to the chiral current in the Skyrme
model. The baryons in the SS-model is realized as a soliton, which at
a time slice is approximately the BPST instanton with a fixed size. We
considered the chiral current in the collective coordinate
quantization of the baryon solution within the large $\lambda$
approximation, and calculated electric charge
distributions (fig.\ \ref{fig:rhoem}), charge radius
\eqref{eq:<r^2>_I=0}, magnetic moments [\eqref{eq:g_I=1=} and
\eqref{eq:g_I=0=1}], and the axial-vector coupling \eqref{eq:g_A=}
of nucleons by taking the masses of the nucleon and $\Delta(1232)$ as
inputs. For most of these quantities, the obtained numerical values
in the SS-model are better close to the experimental values than
in the Skyrme model \cite{ANW}.
We emphasize that in this calculation it is important to quantize the
scale $\rho$ of the baryon solution as an approximate collective
coordinate.
If we treat $\rho$ simply as a constant determined as the minimum of
the potential, agreement with the experimental values becomes worse
for many of the static properties as we mentioned explicitly for the
axial vector coupling in sec.\ \ref{subsec:g_A}.

Although the results of our computation of static properties of
nucleons are phenomenologically rather satisfactory ones,
we have to recall that there remain two problems in our chiral current
$j_{L/R}^\mu(x)$ in the SS-model as we explained in sec.\
\ref{subsec:problemsofj}:
One is the ambiguity of the interpolating function $\gtfn(x,z)$
satisfying the boundary condition \eqref{eq:BCzeta(x,z)}, and the
other is the gauge-noninvariance of the current. As the function
$\gtfn(x,z)$, we took \eqref{eq:ourzeta(x,z)} using the zero-modes
$\psi_\pm(z)$ \eqref{eq:psi_pm} in this paper. However, we have no
convincing reason for choosing this particular $\gtfn(x,z)$.
The choice of $\gtfn(x,z)$ directly affects the results of our
computation. In particular, in the calculation of the axial-vector
coupling in sec.\ \ref{subsec:g_A}, the first term
\eqref{eq:App_psi_A} of the Taylor series at $z=0$ of $\gtfn(x,z)$ for
the axial-vector current was the source of our nice result
\eqref{eq:g_A=}. The gauge-noninvariance is related with the choice of
$\gtfn(x,z)$ as we explained in sec.\ \ref{subsec:problemsofj}.
It is indispensable to resolve these two problems in order to make the
computations carried out in this paper really meaningful.

Reconsideration of another chiral current $\wt{j}_{L/R}^\mu(x)$
\eqref{eq:wtj} obtained from the coupling with the external gauge
fields on the boundaries or from the bulk-boundary correspondence may
also be necessary. The two chiral currents are related by
\eqref{eq:relation}: They are equal to each other up to the EOM and a
trivially conserving term. An advantage of the current
$\wt{j}_{L/R}^\mu(x)$ is that it is free from the two problems of
$j_{L/R}^\mu(x)$ mentioned above. However, $\wt{j}_{L/R}^\mu(x)$  does
not reproduce the chiral current of the Skyrme model in the
low energy limit. Moreover, since the baryon solution in the SS-model
is localized near the origin of the extra fifth dimension, the local
current $\wt{j}_{L/R}^\mu(x)$ itself, which is given in terms of the
field strengths on the boundaries, vanishes for the baryon
configuration.
We saw in sec.\ \ref{subsec:compwtj} that a static property given as
an integration of a quantity containing the chiral current could be
non-vanishing if we take the limit of going to the boundary after
carrying out the integration. This is the case for the isovector
$g$-factor.
It might be that, by taking into account more subtle points in the
current $\wt{j}_{L/R}^\mu(x)$ or by going beyond the large $\lambda$
approximation, we could get nontrivial results from
$\wt{j}_{L/R}^\mu(x)$ for all the static properties of nucleons
including the local charge density.

\section*{Acknowledgements}
We would like to thank H.~Suganuma and S.~Sugimoto for valuable
discussions.
The work of H.~H.\ was supported in part by a Grant-in-Aid for
Scientific Research (C) No.\ 18540266 from the Japan Society for the
Promotion of Science (JSPS).

\appendix

\section{Vector current in the regular gauge}
\label{app:regulargauge}

In sec.\ \ref{sec:staticprop}, we gave an analysis of the static
properties of nucleons based on the baryon solution \eqref{eq:A^cl_SG}
in the singular gauge. Fortunately, the singularity of the solution at
the origin $\xi=0$ did not cause any trouble.
In this appendix, we outline the analysis of the isospin density using
the baryon solution in the regular gauge to explain problems of this
gauge (see sec.\ \ref{subsec:chargedensity} for the isospin density
in the singular gauge).

The baryon solution in the regular gauge is related to the singular
gauge solution via the gauge transformation by $\ginst(\bm{x},z)$.
Concretely, the $SU(2)$ part of the regular gauge solution is given by
\eqref{eq:A^cl_SG} with $\ginst(\bm{x},z)$ and $\ginst(\bm{x},z)^{-1}$
exchanged and $\ol{f}(\xi)$ replaced with $f(\xi)$
\eqref{eq:f(xi)}. The $U(1)$ part of the solution, $\Ah^\cl_M$,
remains unchanged from \eqref{eq:Ah^cl}. The field strengths of the
solution are given by \eqref{eq:F^cl_ij_iz} with
$\ginst^{-1}t_a\ginst$ replaced with $t_a$.
After the introduction of the collective coordinates, the gauge fields
and the field strengths are given by
\eqref{eq:A_i=WA^cl_iW^-1}--\eqref{eq:F_0M} with $\olPhi_a$
\eqref{eq:olPhi} replaced by
\begin{equation}
\Phi_a=\ginst\olPhi_a\ginst^{-1}
=f\bigl(\xi;X^\alpha(t)\bigr)\ginst t_a\ginst^{-1} .
\label{eq:Phi}
\end{equation}

Now, let us consider the five-dimensional isospin density
$\fJ_{V,a}^0(x,z)$ \eqref{eq:J_V,a^0} in the regular gauge.
Instead of the formula \eqref{eq:sum_M[DolPhi,A]=}, here we use the
following one in the regular gauge:
\begin{equation}
\sum_{M=i,z}i\bigl[D^\cl_M\Phi_a,A^\cl_M\bigr]\Bigr|_\regg
=-\frac{8\rho^2\xi^2}{(\xi^2+\rho^2)^3}\,\ginst t_a\ginst^{-1} .
\label{eq:sum_M[DPhi,A]=}
\end{equation}
Since $\ginst t_a\ginst^{-1}$ in \eqref{eq:sum_M[DPhi,A]=}, which is
given by the RHS of \eqref{eq:ginst^-1t_aginst} with the replacement
$x^i\to -x^i$, is not spherically symmetric, we carry out the spatial
angle averaging; $\VEV{x^ax^b}_\Omega=\delta^{ab}\bm{x}^2/3$,
$\VEV{x^i}_\Omega=0$. Then, we get
\begin{equation}
\fJ_{V,a}^0(x,z)\bigr|_\regg
=-\frac{2}{\pi^2}\frac{1}{(\xi^2+\rho^2)^3}
\left(z^2-\frac{\bm{x}^2}{3}\right)I_a ,
\label{eq:J_V,a^0_RG}
\end{equation}
which should be compared with \eqref{eq:J_V,a^0==} in the singular
gauge. Note that the four-dimensional integration $\int\!d^3x\int\!dz$
of \eqref{eq:J_V,a^0_RG} is superficially logarithmically divergent,
and it can take any value depending on the way of integration.
Concretely, this causes the following problem on the isospin density.
Integrating \eqref{eq:J_V,a^0_RG} over $z$, the isospin density in
four dimensions in the regular gauge is given by
\begin{equation}
j_{V,a}^0(t,\bm{x})\bigr|_\regg
=\int_{-\infty}^\infty\! dz\,\fJ_{V,a}^0(x,z)\bigr|_\regg
=-\frac{\rho^2}{4\pi}\frac{1}{(r^2+\rho^2)^{5/2}}\,I_a .
\end{equation}
Integrating this further over $\bm{x}$, we get a conflicting result:
\begin{equation}
\int\! d^3x\,j_{V,a}^0(t,\bm{x})\bigr|_\regg
=-\frac13\,I_a .
\label{eq:intd^3xj_V,a^0_RG=}
\end{equation}
On the other hand, if we carry out the $\bm{x}$ integration first and
then the $z$ integration, we obtain the consistent result:
\begin{equation}
\int_{-\infty}^\infty\!dz\int\!d^3x\,\fJ_{V,a}^0(x,z)\bigr|_\regg
=I_a .
\label{eq:intxintzJ_V,a^0_RG}
\end{equation}
We can show \eqref{eq:intxintzJ_V,a^0_RG} for a generic
$\psi_V(z)$ satisfying the condition $\psi_V(z\to\pm\infty)=1$, not
restricted to $\psi_V(z)=1$ used in the above calculations.
In fact, using that
$D_z\Phi_a\bigr|_\regg=2\rho^2 z/(\xi^2+\rho^2)^2 t_a$
after the angle averaging, we get
\begin{equation}
\int\! d^3x\,\fJ_{V,a}^0(x,z)\bigr|_\regg
=\frac12\drv{}{z}\!\left(\frac{z}{\sqrt{z^2+\rho^2}}\,\psi_V(z)
\right)I_a ,
\end{equation}
and hence \eqref{eq:intxintzJ_V,a^0_RG}.
Note that \eqref{eq:J_V,a^0==} in the singular gauge is of order
$1/\xi^6$ for large $\xi$ and there is no problem of the way of
integrations.

The origin of the above explained trouble in the regular gauge
is the fact that the gauge fields do not vanish sufficiently fast as
$\xi\to\infty$. The baryon solution $A^\cl_M|_\regg$ in the
regular gauge tends to the pure gauge
$-i\ginst\p_M\ginst^{-1}=\cO(1/\xi)$ as $\xi\to\infty$
in contrast with the $\cO(1/\xi^3)$ behavior of the singular gauge
solution \eqref{eq:A^cl_M_SG_explicit}.
This slow falloff of the regular gauge solution would be insufficient
for the condition \eqref{eq:cA_M->0}.

\end{document}